\documentclass[aps,prd,twocolumn,superscriptaddress,preprintnumbers,nofootinbib,longbibliography,floatfix]{revtex4-2}
\usepackage{graphicx}
\usepackage{dcolumn}
\usepackage{bm}
\usepackage{enumitem}
\usepackage{mathtools}
\usepackage{adjustbox}
\usepackage{float}
\restylefloat{table}
\usepackage{subcaption}
\usepackage{multirow}
\usepackage[colorlinks=true
,urlcolor=blue
,anchorcolor=blue
,citecolor=blue
,filecolor=blue
,linkcolor=blue
,menucolor=blue
,pagecolor=blue
]{hyperref}
\usepackage{braket}
\usepackage{amsmath,amsfonts}
\usepackage[table]{xcolor}

\def\begsub#1#2\endsub{\begin{subequations}\label{eq:#1}\begin{align}#2\end{align}\end{subequations}}
\def\ba#1\ea{\begin{align}#1\end{align}}
\def\bas#1\eas{\begin{align*}#1\end{align*}}

\usepackage{acronym}

\begin{document}

%
%

\title{Unsupervised Quantum Circuit Learning in High Energy Physics}
\thanks{This manuscript has been authored by UT-Battelle, LLC, under Contract No. DE-AC0500OR22725 with the U.S. Department of Energy. The United States Government retains and the publisher, by accepting the article for publication, acknowledges that the United States Government retains a non-exclusive, paid-up, irrevocable, world-wide license to publish or reproduce the published form of this manuscript, or allow others to do so, for the United States Government purposes. The Department of Energy will provide public access to these results of federally sponsored research in accordance with the DOE Public Access Plan.}%

\author{Andrea Delgado}
\email{delgadoa@ornl.gov}
\affiliation{Physics Division, Oak Ridge National Laboratory, Oak Ridge, TN 37830}

\author{Kathleen E. Hamilton}
\email{hamiltonke@ornl.gov}
\affiliation{Computer Science and Engineering Division, Oak Ridge National Laboratory, Oak Ridge, TN 37830}

\begin{abstract}
Unsupervised training of generative models is a machine learning task that has many applications in scientific computing.  In this work we evaluate the efficacy of using quantum circuit-based generative models to generate synthetic data of high energy physics processes.  We use non-adversarial, gradient-based training of quantum circuit Born machines to generate joint distributions over 2 and 3 variables.  
\end{abstract}

\maketitle


%
%

\acrodef{HEP}{High-energy physics}
\acrodef{LHC}{Large Hadron Collider}

%
%

\section{Introduction}
\label{sec:intro}

High-energy physics seeks to understand matter at the most fundamental level. Vast and complex accelerators have been built to elucidate the dynamical basis of the fundamental constituents, the quarks, and the gluons. At these large-scale facilities, high-performance data storage and processing systems are needed to store, access, retrieve, distribute, and process experimental data. Experiments like the Compact Muon Solenoid (CMS) at the Large Hadron Collider (LHC) are incredibly complex, involving thousands of detector elements that produce raw experimental data rates over a Tb/sec, resulting in the annual productions of datasets in the scale of hundreds of Terabytes to Petabytes. In addition, the manipulation of these complex datasets into summaries suitable for the extraction of physics parameters and model comparison is a time-consuming and challenging task.

A crucial element of any analysis workflow in particle physics involves simulating the physical processes and interactions at these facilities to develop new theories and models, explain experimental data, and characterize background. These simulations also allow for studying detector response and plan detector upgrades. The simulation of particle physics interactions is often computationally intensive, taking up a significant fraction of the computational resources available to physicists.

Recently, alternative methods for detector simulation and data analysis tasks have been explored, like machine learning applications and quantum information science (QIS). QIS is a rapidly developing field focused on understanding the analysis, processing, and transmission of information using quantum mechanical principles and computational techniques. QIS can address the conventional computing gap associated with HEP-related problems, specifically those computational tasks that challenge CPUs and GPUs, such as efficient and accurate event generators. Thus, the relevance of having an efficient simulation mechanism that can faithfully reproduce particle interactions after a high-energy collision has sparked the development of alternative methods. One particular example is the use of generative models such as generative adversarial networks (GANs) \cite{originalGANs}, which have been utilized in HEP as a tool for fast Monte Carlo (MC) simulations \cite{CaloGAN, deOliveira2017, Musella2018}, and as a machine learning-enhanced method for event generation \cite{Butter2019, Disipio2019Dijetgan, hashemi2019lhc, Alanazi2021}. Some of these studies report up to five orders of magnitude decrease in computing time. A crucial feature of generative models is their ability to generate synthetic data by learning from actual samples without knowing the underlying physical laws of the original system. In some studies, generative models have been shown to overcome the statistical limitations of an input sample in subtraction of negative-weight events in samples generated beyond leading-order in QCD \cite{Butter2019}, and to increase the statistics of centrally produced Monte Carlo datasets \cite{hashemi2019lhc}. Quantum-assisted models have also been proposed for Monte Carlo event generation \cite{bravoprieto2021stylebased}, detector simulation \cite{Chang2021, chang2021quantum}, and determining the parton distribution functions in a proton \cite{perez2021determining}.
In this work, we successfully trained a quantum generative model to reproduce kinematic distributions of particles in \textit{pp} interactions at the LHC, with high fidelity. Quantum circuit Born machines (QCBM) trained via gradient-based optimization \cite{liu2018differentiable,hamilton2019generative,benedetti2018generative} are examples of circuit-based parameterized models that can be trained on near-term quantum platforms. 

\section{Model and learning algorithm}
\label{sec:model}
\begin{figure}
    \centering
    \includegraphics[width=.98\linewidth]{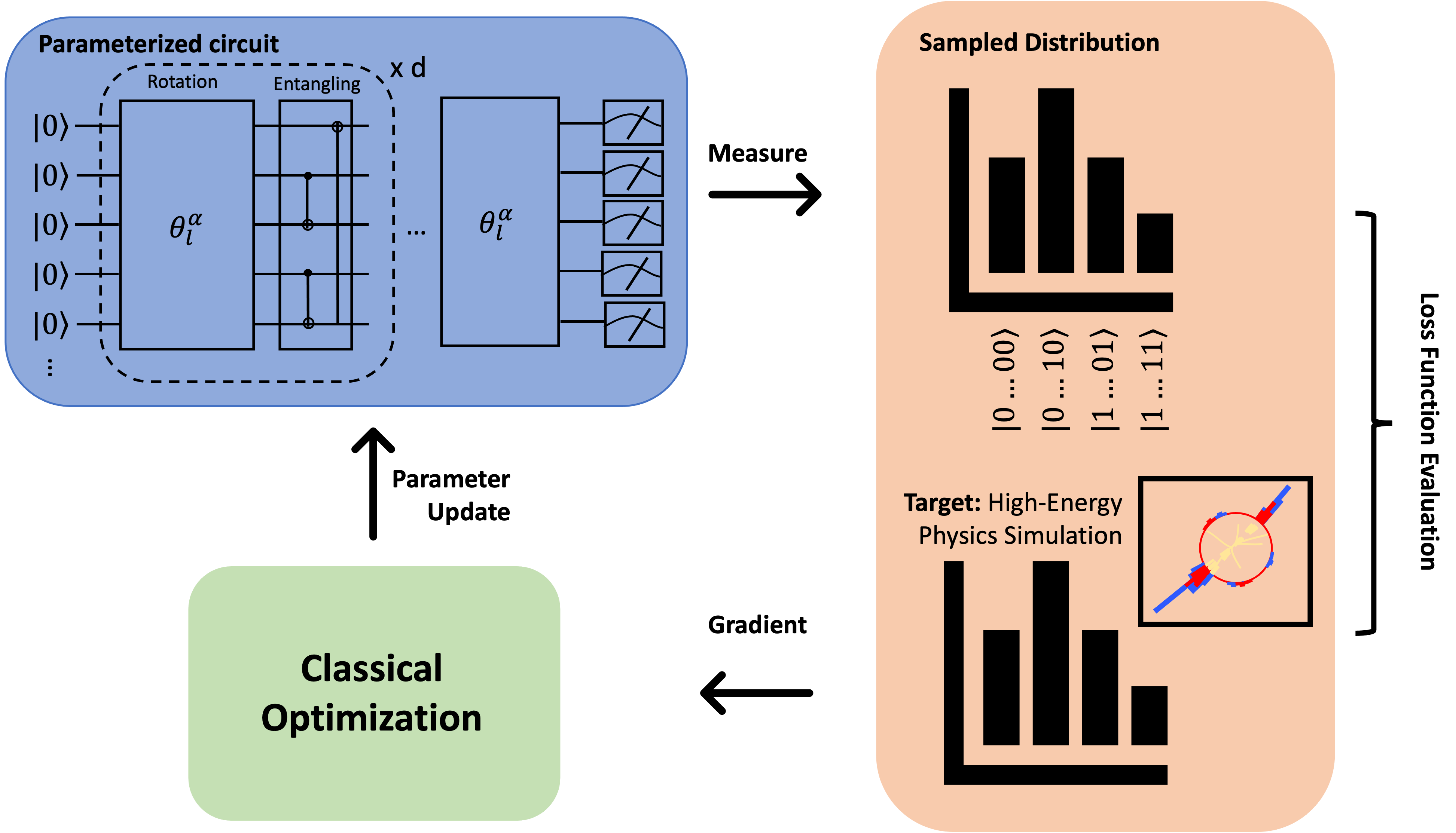}
    \caption{ Diagram of the differentiable QCBM training scheme.}
    \label{fig:qcbmmodel}
\end{figure}

Although generative models trained in adversarial settings have been proven to be a valuable tool in HEP, for this work, we focus on generative models trained with non-adversarial methods. Expressly, we set up and train multiple QCBM using data-driven circuit learning (DDCL) \cite{benedetti2018generative,liu2018differentiable,hamilton2019generative}. DDCL employs a classical-quantum hybrid feedback loop, as described in Fig. \ref{fig:qcbmmodel}.

Generating synthetic data for this HEP application consists of three stages: first, we encode our data of of $M$ observations with $N$-dimensional, numerical features ($\mathcal{D} = \{ X^{(1)}, \dots, X^{(m)}, \dots, X^{(M)}\}$) into a distribution $P(x)$ defined over finite length bitstrings ($x$); second, we train a parameterized circuit ansatz using DDCL to prepare an approximation of this distribution $\widetilde{P}(x)$; third, we generate synthetic data by decoding bitstrings sampled from $\widetilde{P}(x)$. In the following subsection, we describe these stages in greater detail.

\subsection{Data Encoding}
\label{sec:encoding}
The kinematic distributions of a particle jet in a high-energy collider experiment such as the LHC can be constructed as a 2-dimensional joint distribution (over $(p_T,m)$ features) or 3-dimensional joint distributions (over $(p_T,m,\eta)$ features). For a QCBM constructed with $Q$ total qubits, $P(x)$ is constructed by concatentating lists of binary bitstrings which encode the marginals of individual features in a classical dataset. The N-dimensional correlated data features  ($X^{(m)}$) are encoded as $2^{Q}$-length binary strings ($x_i$).  

Individual features are encoded in subsets of $q = Q/N$ qubits by discretizing the marginal distribution into $2^{q}$ bins and each bin index is converted to a $2^q$-length binary bitstring. The final $2^Q$-length bistrings are constructed by concatenating the N feature $2^q$-length binary bitstrings and normalizing the amplitudes.

\subsection{Quantum Circuit Model and Training}
\label{sec:qcbm}
The QCBM is an example of an implicit model for generative learning \cite{mohamed2016learning} that generates data by measuring the system as a Born machine. A QCBM is a parameterized unitary $\mathcal{U}(\Theta)$ that prepares $Q$-qubits in the state $\ket{\psi_{\Theta}}=\mathcal{U}(\Theta)\ket{\psi_0}$. The initial state $\ket{\psi_0}$ is fixed, and in this study, we use:  the all zero-state $\ket{\psi_0}= \ket{0}^{\otimes Q}$; a product state of $Q/2$ Bell states $\ket{\Phi^+}^{\otimes Q/2}$; and a product state of $Q/3$ GHZ states $\ket{\psi_0}= \ket{\mathrm{GHZ}}^{\otimes Q/3}$.

Measuring $\ket{\psi_{\Theta}}$ in a fixed basis $\mathcal{M}$ \footnote{We use the Z-basis $\mathcal{M} = Z_i \otimes \dots \otimes Z_Q$ unless otherwise noted.} requires sampling from the state with $N_{shots}$ shots. This defines a classical distribution over the $2^Q$ computational basis states $\widetilde{P}_{\Theta}(x)$ that is used in training, and later used to generate the synthetic data $\widetilde{X}$. 

\subsubsection{Parameterized Quantum Circuit}
Finding the ideal unitary $\mathcal{U}(\Theta)$ is dependent on the parameterized quantum circuit (PQC). The PQC design plays an essential role in the performance of many variational hybrid quantum-classical algorithms \cite{sim2019expressibility} by defining the hypothesis class. For our application, PQCs must be able to model different types of correlations in the input data. This requires circuits to prepare strongly entangled quantum states and the ability to explore Hilbert space. 

Variational algorithms have been implemented using quantum circuits composed of a network of single and two-qubit operations, with rotation angles serving as variational parameters. The pattern defining the network of gates is referred to as a \textit{unit-cell} or \textit{circuit block} that can be repeated to suit the needs of the application or task at hand. Recently, the term Multilayer Quantum Circuit (MPQC) was coined to describe this type of variational circuit architecture \cite{du2020expressive}. In this study, we train two circuit templates or ansatz. Each circuit is defined using a $Q$-qubit register and specified by the number of layers $d$, consisting of a rotation and an entangling component. Ansatz 1 has a combined rotation and entangling gates in a ``Brick Layer'' or ``Simplified 2-design'' architecture and has been shown to exhibit important properties to study barren plateaus in quantum optimization landscapes \cite{mcclean2018barren,cerezo2021cost}. Ansatz 2 was chosen due to the low correlation displayed between variables and is an extension of ansatz employed in benchmarking tasks \cite{hamilton2019generative}. 

\begin{figure}
    \centering
    \includegraphics[width=\linewidth]{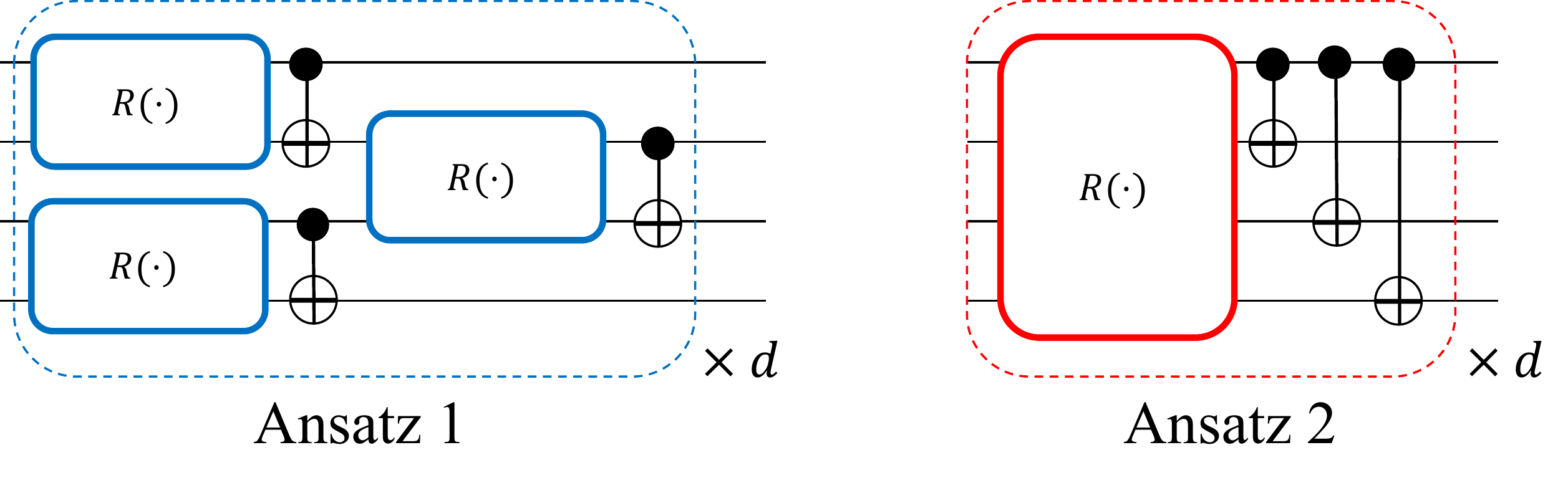}
    \caption{ Diagram for one layer of the circuit templates used to construct and train the QCBM. Both ansatz are constructed using a layered pattern of rotation and entangling gates. For both ansatz a final layer of rotation gates is added before measurement.}
    \label{fig:circuid_diagram}
\end{figure}

The diagram for one layer of each template is shown in Fig. \ref{fig:circuid_diagram}. The rotation gate layers are parameterized by the arbitrary single-qubit rotation gate  implemented in PennyLane \cite{bergholm2020pennylane}, which has 3 rotation angles $R(\Theta)_{\ell}^{(i)}=R(\omega,\theta,\phi)_{\ell}^{(i)}$, where the layer index $\ell$ runs from 0 to $d$, and $i$ is the qubit index. 

We choose to train two ansatz that can be embedded onto near-term NISQ devices: either using a 1D chain of qubits (for Anasatz 1), or the "disconnected tree" configuration (for Ansatz 2). For Ansatz 1 used in this study, the number of parameters to optimize during training is $N_{\rm par}(N_{\ell},Q)=3N_{\ell}[2Q - 1(2) ] + 3Q$ for QCBM with odd(even) qubits $Q$.
For Ansatz 2, $N_{\rm par}(N_{\ell},Q)=3(N_{\ell}+1)Q$, regardless of whether $Q$ is even or odd. 
\captionsetup[figure]{textfont=normalfont,justification=raggedright}
\subsubsection{Training and cost function}
Training a QCBM via DDCL optimizes $\Theta$ by minimizing a loss function $\mathcal{L}(P,\widetilde{P}_{\Theta})$ such that $\widetilde{P}_{\Theta}(x) \approx P(x)$. We use the Jensen-Shannon (JS) divergence- a differentiable function that compares two distributions $P(x)$ and $\widetilde{P}_{\Theta}(x)$:
\begin{equation}
    JS(P|\widetilde{P}_{\Theta}) =
    \frac{1}{2}\sum_{x}\left[ P\log{\left(\frac{P}{M}\right)} + \widetilde{P}_{\Theta}\log{\left(\frac{\widetilde{P}_{\Theta}}{M}\right)}\right]
\label{eq:jsloss}
\end{equation}
where $M = (P + \widetilde{P}_{\Theta})/2$. The minimum value of the loss function $JS(P|\widetilde{P}_{\Theta})= 0$ is achieved when $\widetilde{P}_{\Theta}(x) = P(x)$. The model parameters $\Theta$ are optimized using classical gradient descent methods so that the loss function is minimized.
The gradient of the loss function is computed with respect to the circuit parameters using the parameter shift rule \cite{schuld2019evaluating}.  Each QCBM is trained using the Adam gradient-based optimizer \cite{kingma2014adam} available in the PennyLane library \cite{bergholm2020pennylane}. 

\subsection{Measurement decoding and post-processing}
The output of a QCBM is $\widetilde{P}_{\Theta}(x)$: a classical distribution over $2^Q$-length binary strings.  To convert $\widetilde{P}_{\Theta}(x)$ to numerical n-dimensional features ($\widetilde{\mathcal{D}} = \{ \widetilde{X}^{(1)},\dots,\widetilde{X}^{(m)},\dots  \}$), we reverse the steps described in Section \ref{sec:encoding}: each binary string $2^Q$ length string is disassociated into $N$ composite strings each of length $2^q$, and a float value randomly drawn from a uniform distribution defined with the bin edges previously used to map the samples $x_{m}$ into the binary basis to generate $P(x)$. 

\section{Validation on LHC dataset}
\label{sec:dataset}
One of the big computational challenges in HEP is the considerable computing time required to model the behavior of subatomic particles both at the vertex and detector level. In Section \ref{sec:model}, we introduce the architecture of a quantum generative model that aims to provide an alternative to traditional Monte Carlo (MC) methods in the context of data augmentation. Thus, to validate the proposed model, we consider the simulation of the production of pairs of jets in $pp$ interactions at the LHC. The dataset \cite{Disipio2019Dijetgan} consists of 10 million di-jet events generated using \textsc{MadGraph5} \cite{Alwall:2011uj} and \textsc{PYTHIA8} \cite{Sjstrand2008}, corresponding to a center of mas energy of 13 TeV and an integrated luminosity of about 0.5 $fb^{-1}$. The response of the detector was simulated by a \textsc{DELPHES} \cite{deFavereau2014} fast simulation, using settings that resemble the ATLAS detector. An average of 25 additional soft-QCD $pp$ collisions (pile-up) were added to the simulation to mimic the conditions of a typical collider event realistically. Jets were reconstructed using the anti-$k_{T}$ \cite{Cacciari2008} algorithm as implemented in \textsc{FastJet} \cite{cacciari2006fastjet}, with a distance parameter $R=1.0$. Selection cuts on $p_{T}$  and $H_{T}$ were applied, reducing the sample size to about 4 million events. The kinematic distributions of the leading jet on the di-jet system are used to validate the QCBM models and evaluate its performance in a real-world application.

\section{Results}
In this section, we report on the results of training a QCBM to prepare the target distribution of the dataset described in Section \ref{sec:dataset}. The model performance is evaluated by studying the JS divergence value throughout the training and comparing target (MC expectation) and generated (the output of the trained QCBM model) distributions. We explore the encoding of the target distributions for 2(3) variable joint distributions into 8(12) qubit systems. This scheme allows for a four-qubit encoding per distribution. Unless noted, each marginal distribution is encoded in a target state binned over $2^{4} = 16$ basis states. We trained the QCBM circuits in the absence of noise to obtain a set of optimal parameters $\Theta$. Then, the circuits were deployed using the trained parameters on IBM quantum devices to study the effect of noise in the loss landscape, reproducing the target distribution. Finally, a local parameter tuning scheme was applied to improve the performance in the presence of noise.

\subsection{Training with noiseless qubits}
\label{sec:training}
\subsubsection{2D Distributions}
\label{sec:2D_correlated_distributions}
\begin{figure}[htbp]
\centering
\includegraphics[width=\linewidth]{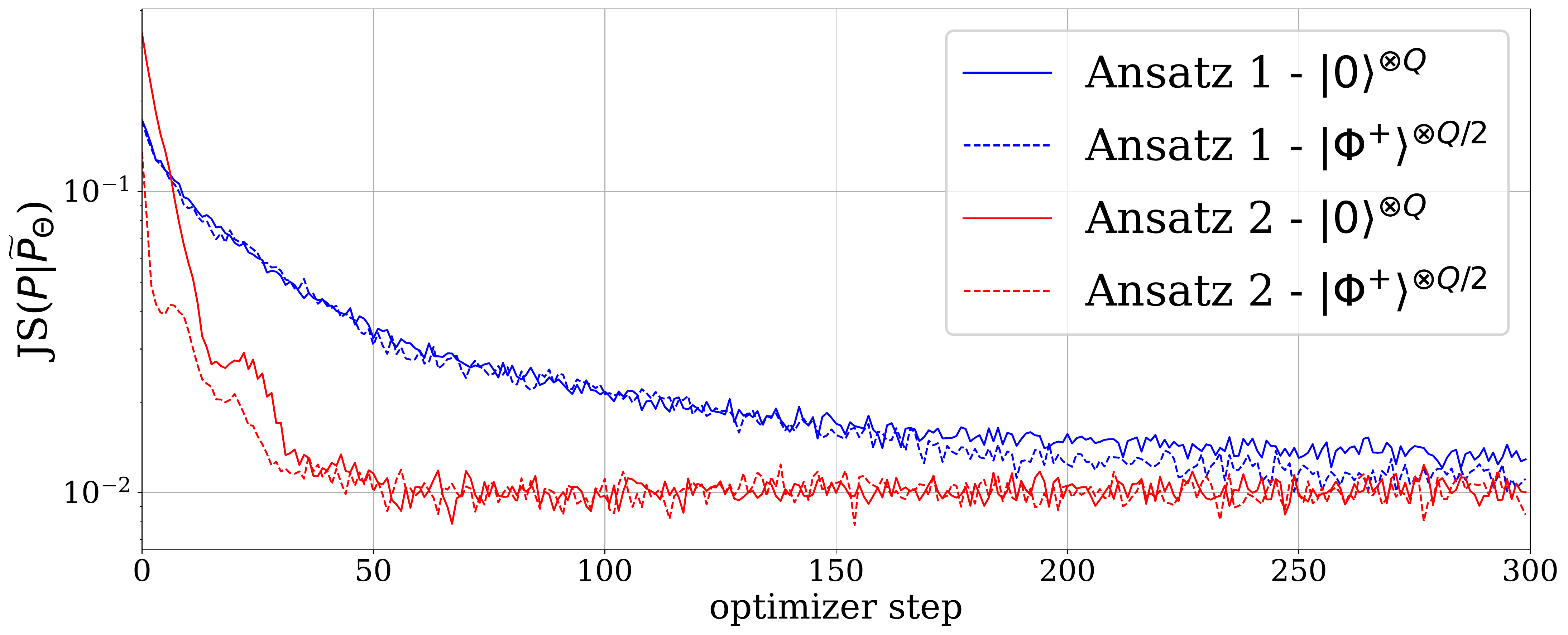}
\caption{JS divergence value as a function of training step using $N_{shots}=8192$. Blue(red) points correspond to Ansatz 1(2). Circuits were initialized in the all zero state (solid lines) or four Bell states (dashed lines) and trained to learn a 2D joint distribution.}
\label{fig:2D_loss}
\end{figure}

In Figure \ref{fig:2D_loss}, the JS divergence values are plotted as a function of training step. The circuits were trained using $N_{shots}=8192$ to prepare a joint 2D distribution corresponding to the marginal distributions of the leading jet transverse momentum ($p_{T}$) and mass, both binned over the 16 basis states corresponding to four qubits. Circuits were constructed using the ansatz configurations shown in Figure \ref{fig:circuid_diagram} and trained to start from either the all-zero state ($\ket{\psi_0}= \ket{0}^{\otimes 8}$), or from a product of four Bell states ($\ket{\Phi^+}^{\otimes 4}$). We fixed the number of layers $N_{layers}=6$ for Ansatz 1 (blue) and 12 for Ansatz 2(red). Each circuit is trained for 300 steps of Adam with a learning rate $\alpha=0.01$.  

\captionsetup[figure]{textfont=normalfont,justification=raggedright}

\begin{figure}[htbp]
    \centering
    \includegraphics[width=0.95\linewidth]{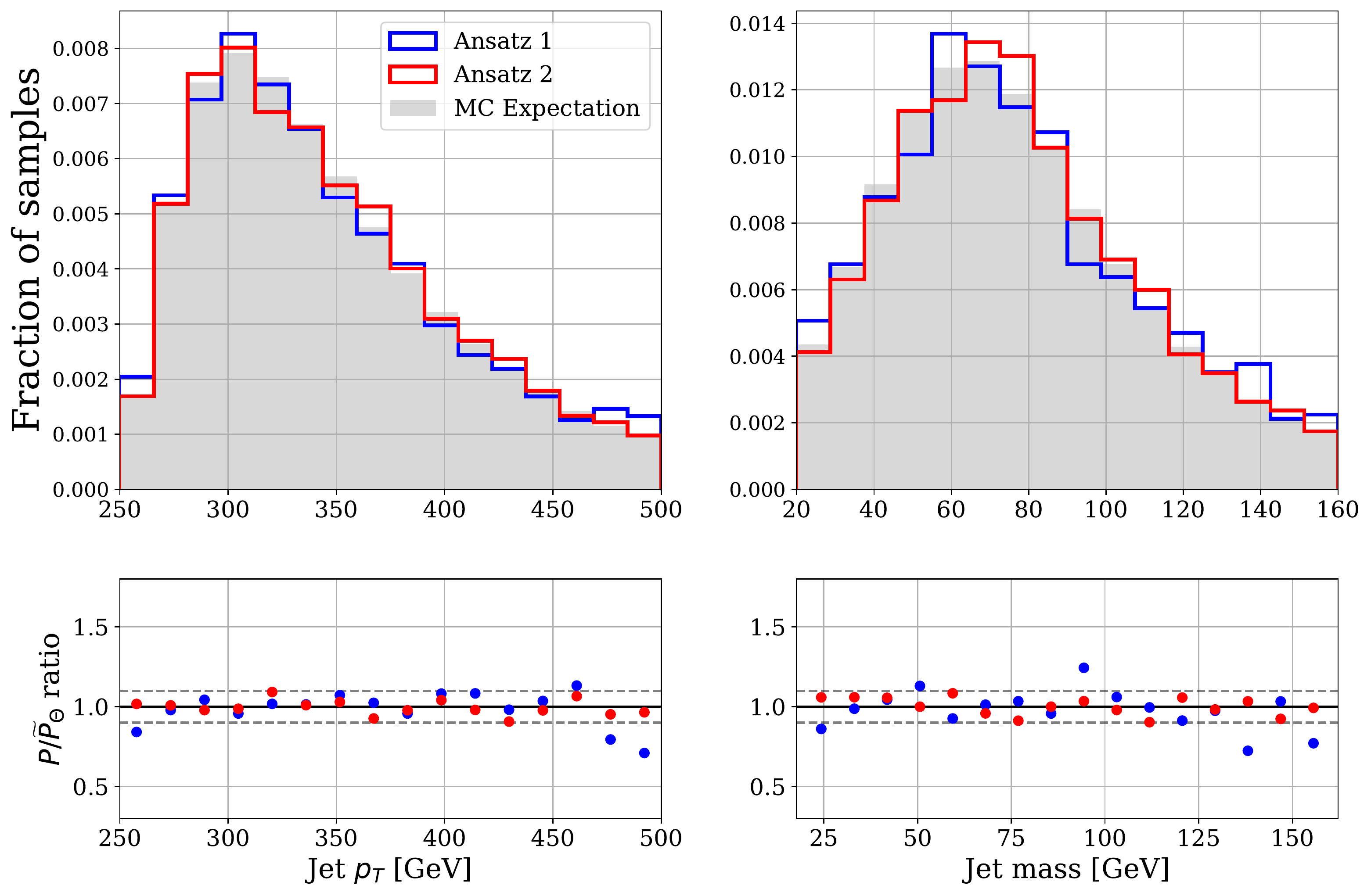}
    \caption{\textbf{Top:} Target and sampled distributions. Blue(red) points correspond to Ansatz 1(2). Circuits were initialized in the all-zero state ($\ket{\psi_0}= \ket{0}^{\otimes 8}$) and trained to learn a 2D joint distribution. 
    \textbf{Bottom:} Ratio of target and sampled distributions with horizontal guide lines marking $P/\widetilde{P}_{\Theta}=0.9$ and $P/\widetilde{P}_{\Theta}=1.1$.}
    \label{fig:2D_allzero}
\end{figure}

From Figure \ref{fig:2D_loss} we can conclude that: Ansatz 2 converges to a stable JS divergence value much faster than Ansatz 1, and the training of the QCBM is not affected by the choice of the initial state. Figure \ref{fig:2D_allzero} displays the distributions of samples generated via projective measurements on the qubits in the trained circuits. We observe that the data generated resembles the target distributions with high fidelity, with a slightly better agreement for data generated by sampling from the QCBM constructed with Ansatz 2 (red).

\begin{figure}[htbp]
\centering
\includegraphics[width=0.95\linewidth]{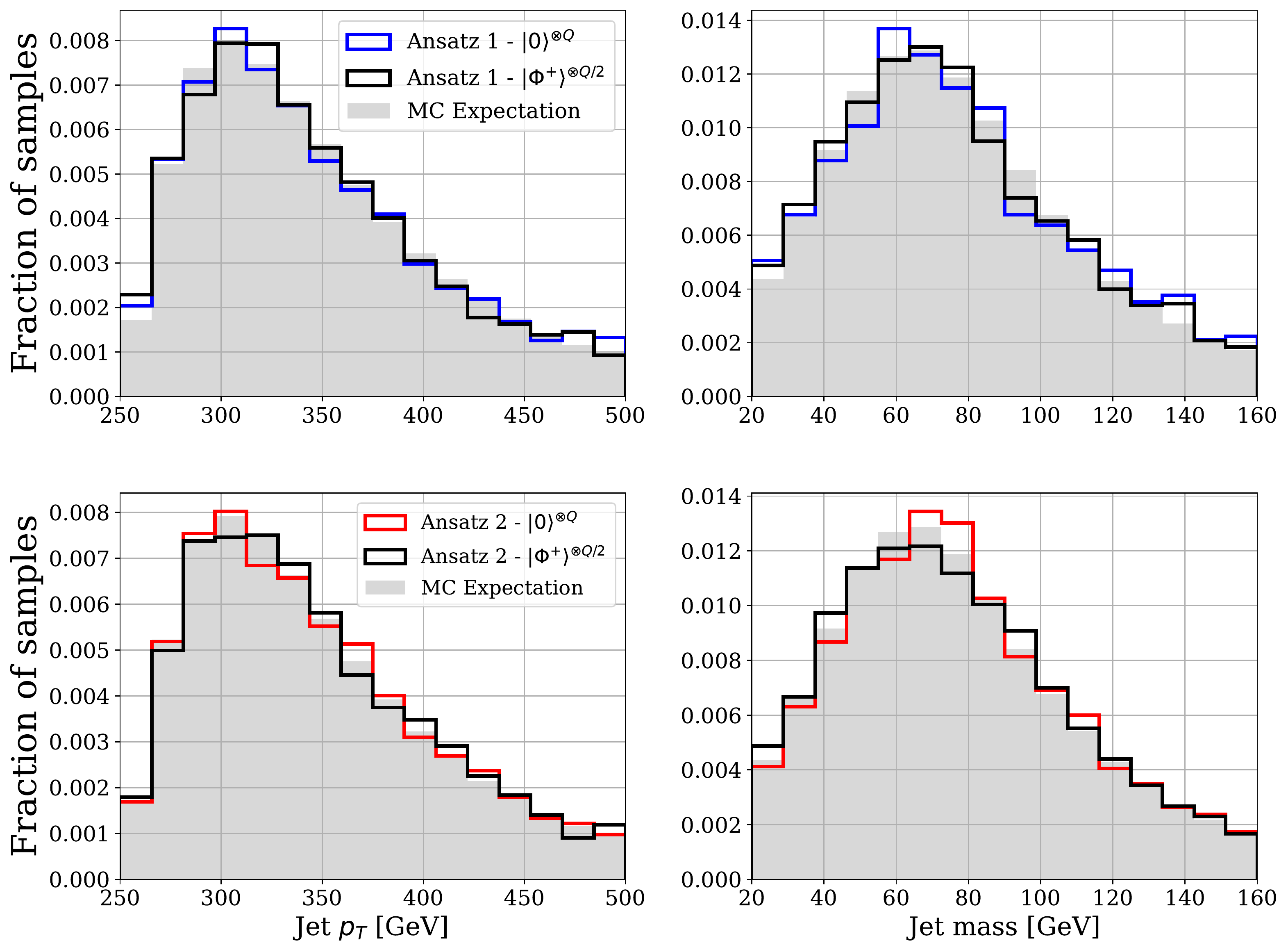}
\caption{\textbf{Top:} Target and sampled distributions constructed using Ansatz 1. Circuits were initialized in the all-zero state (blue) and four Bell states (black). \textbf{Bottom:} Target and sampled distributions constructed using Ansatz 2. Circuits were initialized in the all zero-state (red) and four Bell states (black).}
\label{fig:2D_all}
\end{figure}

In Figure \ref{fig:2D_all}, we compare the sampled and target distributions obtained when starting the training from either an all-zero state ($\ket{\psi_0}= \ket{0}^{\otimes 8}$) or from a product of four Bell states ($\ket{\Phi^+}^{\otimes 4}$). We define a similarity measure to perform a systematic comparison of the marginal distributions for each feature by computing the mean absolute error (MAE) per bin between all normalized target and sampled marginal distributions:
\begin{equation}
    D(p|\widetilde{p}(\Theta)) = \frac{1}{N2^q}\sum_n^{N}\sum_{i}^{2^q} \left|p^{(n)}_{i} - \widetilde{p}_{i}^{(n)}(\Theta)\right|.
    \label{eq:dif}
\end{equation}

\captionsetup[figure]{textfont=normalfont,justification=raggedright}

From both Figure \ref{fig:2D_all} and Table \ref{table:2dcomp}, we can conclude that Ansatz 2, initialized in the all-zero state, reproduces the feature marginals with the highest fidelity. On the other hand, the difference in $D(p|\widetilde{p}(\Theta))$ for the two initial configurations considered is negligible, considering a statistical error proportional to $1/\sqrt{N_{samples}}\sim 0.0005728$, implying that the training is independent of circuit initialization.

\captionsetup[figure]{textfont=normalfont,justification=raggedright}
\begin{figure}[htbp]
    \centering
    \includegraphics[width=0.95\linewidth]{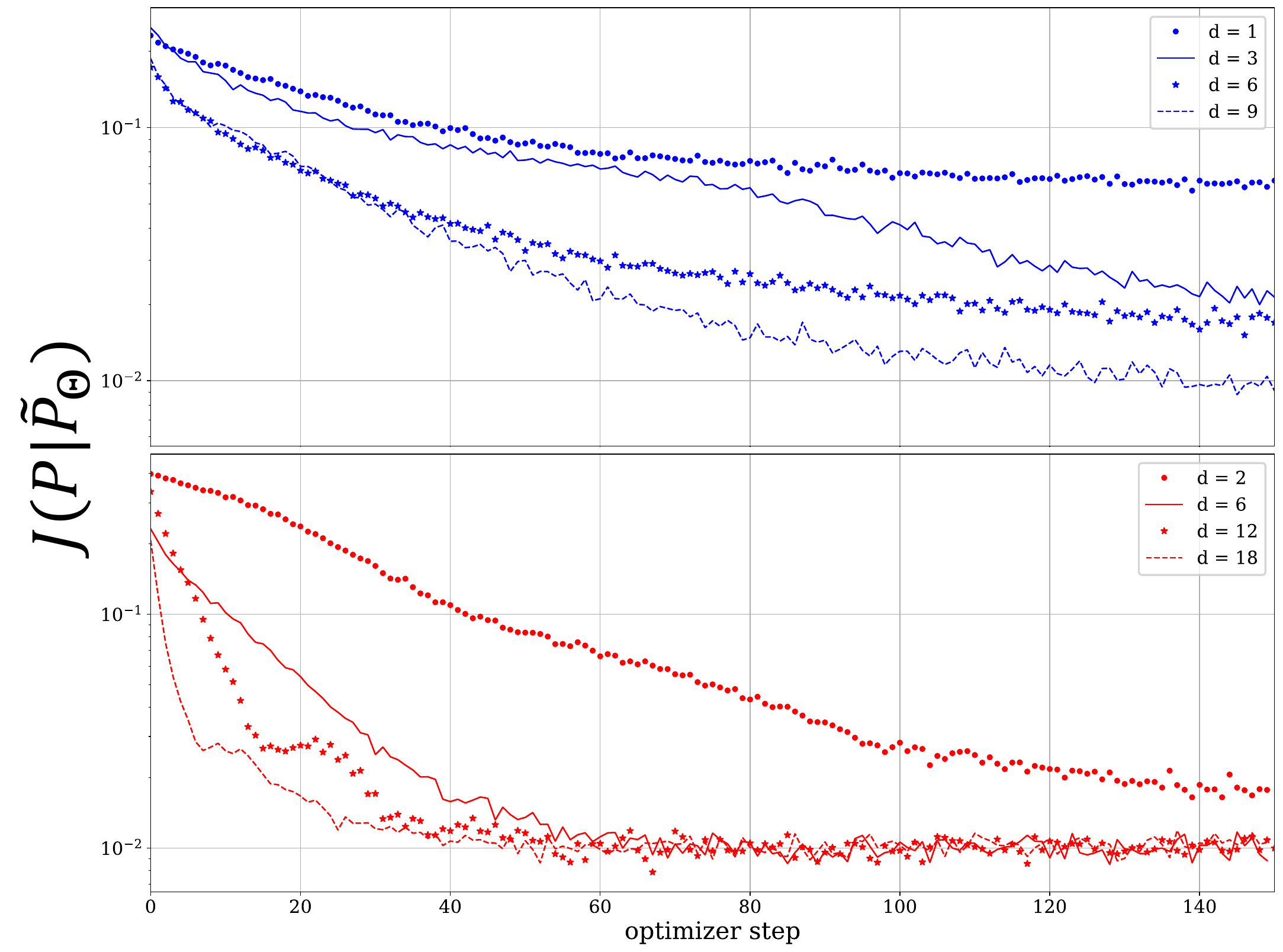}
    \caption{Loss as a function of training step using $N_{shots}=8192$. The different curves represent a different number of layers $d$, used to construct the 8 qubit QCBM. \textbf{Top:} Ansatz 1 (blue), and \textbf{Bottom:} Ansatz 2 (red).}
    \label{fig:2D_loss_nlayer}
\end{figure}

\begin{table}[htbp]
\centering
\renewcommand{\arraystretch}{1.5}
 \begin{tabular}{|c| c |  p{2.5cm}|c |} 
 \hline
 Ansatz & $N_{\rm par}$ & Initial State & $D(p|\widetilde{p}(\Theta))$  \\
 \hline\hline
 1 & \multirow{2}{*}{276} & $\ket{\psi_0} = \ket{0}^{\otimes 8}$ & 0.007153 \\ 
 1 & &  $\ket{\psi_0} =\ket{\Phi^+}^{\otimes 4}$     & 0.006767 \\
 \hline
 2 &  \multirow{2}{*}{312} & $\ket{\psi_0}= \ket{0}^{\otimes 8}$ & 0.005452 \\
 2 &   &  $\ket{\psi_0} =\ket{\Phi^+}^{\otimes 4}$ & 0.005696 \\ 
 \hline
\end{tabular}
\caption{$D(p|\widetilde{p}(\Theta))$ for 8 qubit QCBM.}
\label{table:2dcomp} 
\end{table}

Another important factor in the process of building the circuit to prepare the target state is the number of layers the template or ansatz in Figure \ref{fig:circuid_diagram} is repeated. This choice will also determine the number of trainable parameters in the model. In Figure \ref{fig:2D_loss_nlayer}, the JS divergence value is plotted as a function of training steps for $d=\{1,3,6,9\}$ layers in Ansatz 1; and $d=\{2,6,12,18\}$ layers in Ansatz 2.  We can see from this plot that the number of layers used to construct Ansatz 2 has little impact on the minimal JS value reached after the training converged. Nonetheless, it affects how fast the model reaches this minimal JS value. On the other hand, the training performance of Ansatz 1 is highly dependent on the number of layers used to construct the circuit. We chose to use $d=6(12)$ to construct our QCBM with Ansatz 1(2) to keep an optimal balance between training time and performance.
\captionsetup[figure]{textfont=normalfont,justification=raggedright}
\begin{figure}
    \centering
    \includegraphics[width=\columnwidth]{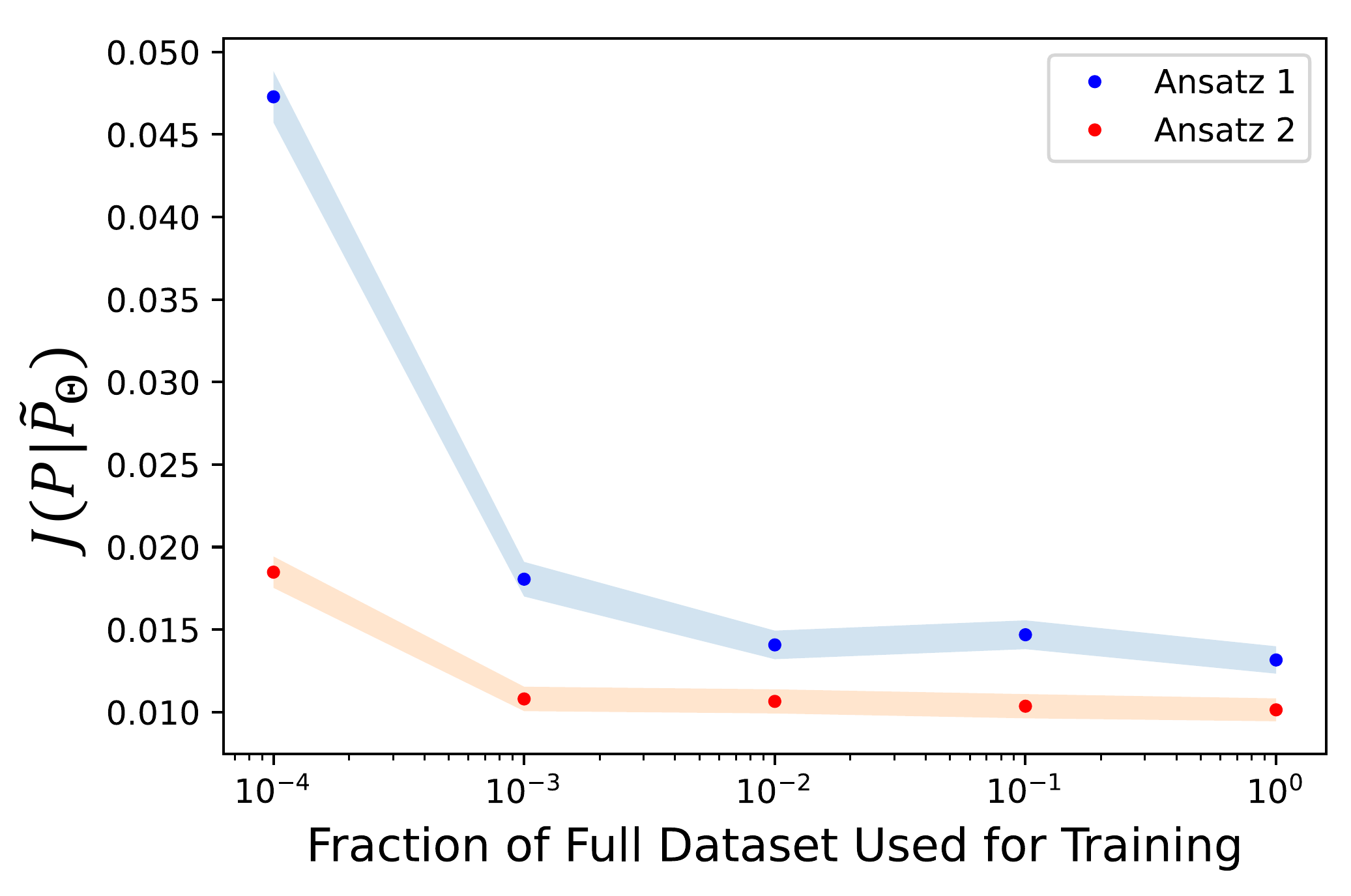}
    \caption{Mean loss between the complete dataset and the distribution generated by 8-qubit QCBM models QCBM circuits plotted as solid circles in blue(red) for Ansatz 1(2). Each QCBM model was trained on partial data. Shaded regions correspond to mean JS divergence value $\pm \sigma$ (one standard deviation).}
    \label{fig:data_fraction}
\end{figure}

 This study proposes using quantum generative models as a data augmentation tool. The results presented in this section thus far were obtained by training QCBMs on a target distribution using close to 4 million events, which is the typical number for any analysis campaign in HEP. We also investigated how reducing the training dataset size affects the trained model's fidelity to reproduce the target distribution. In Figure \ref{fig:data_fraction}, the horizontal axis represents the fraction of the initial training dataset of 4 million events used to train the QCBM. Once the model is trained, using a fraction of the full dataset, the JS divergence metric is evaluated on the target distribution generated using the whole dataset. The distribution was obtained by evaluating the QCBM with the trained parameters with 8192 shots. The sampling process was repeated 1000 times, and the mean is reported in Figure \ref{fig:data_fraction} as a solid blue(red) dot for Ansatz 1(2). The bands correspond the mean JS divergence value $\pm \sigma$ (one standard deviation).
\begin{table}[H]
\centering
\scriptsize
\begin{subtable}[t]{0.5\textwidth}
\centering 
\caption{Monte Carlo (Ground Truth)}
        \label{tab:MC_2d_correlation}
\resizebox{0.5\textwidth}{!}{
\begin{tabular}{|p{1cm}|c|c|}
\cline{2-3}
\multicolumn{1}{c|}{} & $p_{T}$ & mass \\
\hline
 $p_{T}$ & {\cellcolor{gray!20}-} & 0.2 \\ \hline
 mass & 0.2 & {\cellcolor{gray!20}-}  \\ \hline
\end{tabular}
}
\end{subtable}
\begin{subtable}[t]{0.5\textwidth}
\centering
\caption{Ansatz 1}
\label{tab:ansatz1_2d_correlation}
\resizebox{0.7\textwidth}{!}{
 \begin{tabular}{| p{1cm} | c | c | c |c |}
\cline{2-5}
 \multicolumn{1}{c|}{} & \multicolumn{2}{c|}{$p_{T}$} & \multicolumn{2}{c|}{mass} \\ 
\cline{2-5}
 \multicolumn{1}{c|}{} & $\ket{0}^{\otimes 8}$  & $\ket{\Phi^+}^{\otimes 4}$ &  $ \ket{0}^{\otimes 8}$  & $\ket{\Phi^+}^{\otimes 4}$ \\ [0.5ex] 
 \hline
 $p_{T}$ & \multicolumn{2}{c|}{\cellcolor{gray!20}-} & {\cellcolor{olive!20} 0.19} & {\cellcolor{teal!20} 0.12} \\
 \hline
 mass    & {\cellcolor{olive!20} 0.19} & {\cellcolor{teal!20} 0.12} & \multicolumn{2}{c|}{\cellcolor{gray!20}-} \\
 \hline
 \end{tabular}
 }
 \end{subtable}
 \begin{subtable}[t]{0.5\textwidth}
\centering
\caption{Ansatz 2}
\label{tab:ansatz2_2d_correlation}
\resizebox{0.7\textwidth}{!}{
 \begin{tabular}{| p{1cm} | c | c | c |c |}
\cline{2-5}
 \multicolumn{1}{c|}{} & \multicolumn{2}{c|}{$p_{T}$} & \multicolumn{2}{c|}{mass}  \\ 
\cline{2-5}
 \multicolumn{1}{c|}{} & $\ket{0}^{\otimes 8}$  & $\ket{\Phi^+}^{\otimes 4}$ & $ \ket{0}^{\otimes 8}$  & $\ket{\Phi^+}^{\otimes 4}$ \\ [0.5ex] 
 \hline
 $p_{T}$ & \multicolumn{2}{c|}{\cellcolor{gray!20}-} & {\cellcolor{olive!20} -1.0e-3} & {\cellcolor{teal!20} -9.1e-3}  \\
 \hline
 mass    & {\cellcolor{olive!20} -1.0e-3} & {\cellcolor{teal!20} -9.1e-3} &  \multicolumn{2}{c|}{\cellcolor{gray!20}-} \\
 \hline
 \end{tabular}
 }
 \end{subtable}
 \caption{Correlation matrices between jet $p_{T}$ and mass (m) variables in the \textbf{(a)} target distribution, and samples obtained from the evaluation of the QCBMs constructed using \textbf{(b)} Ansatz 1 and \textbf{(c)} Ansatz 2 in Figure \ref{fig:circuid_diagram} with the trained parameters. Values displayed for initial states prepared in the all-zero state (olive) and a product of 4 Bell states (teal).}
 \label{table:2dcorrMatrix}
\end{table}

Finally, we report on the correlation matrix between the jet $p_{T}$ and mass variables used to construct the target distribution. In \ref{tab:MC_2d_correlation}, the values associated with the target distribution are displayed in black. If the trained QCBM learned the joint distribution, one would expect to recover the correlation matrix when evaluating the QCBM with the trained parameters. The results in Table \ref{table:2dcorrMatrix} indicate that this is true for the QCBM constructed with Ansatz 1 (red), but not for the QCBM constructed with Ansatz 2. The correlation matrix for the latter case indicates that there is little correlation between the marginal distributions in the synthetic samples.

\subsubsection{3D Distributions}
\label{sec:3D_correlated_dist}
\begin{figure}
    \centering
    \includegraphics[width=\columnwidth]{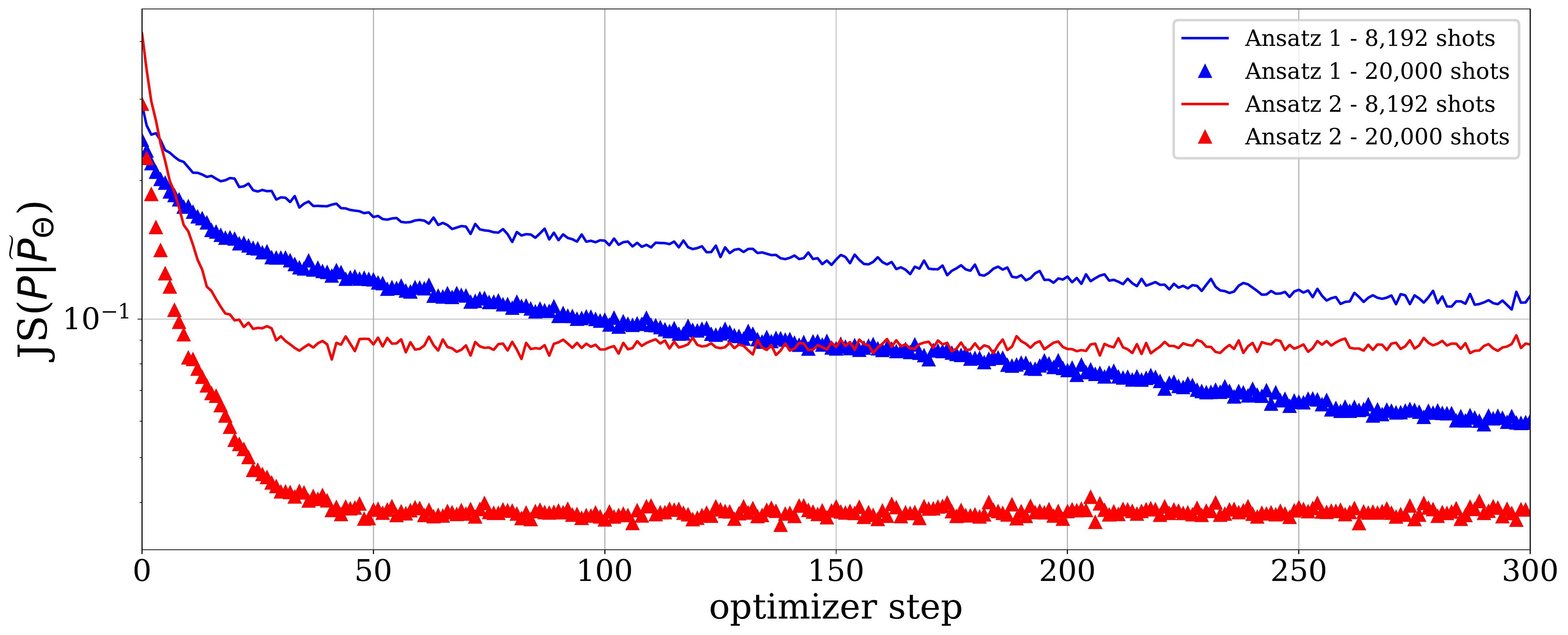}
    \caption{JS divergence as a function of training step using $N_{shots}=8192$ (solid lines) and $N_{shots}=20,000$ (triangles). Blue(red) points correspond to Ansatz 1(2). Circuits were initialized in the all zero state and trained to learn a 3D joint distribution. QCBM were initialized in the all-zero state ($\ket{\psi_0}= \ket{0}^{\otimes 12}$.
    }
    \label{fig:3D_loss}
\end{figure}

To understand how the trainability of non-adversarial generative models scales with the number of quantum registers, we increased the number of qubits in our model from 8 to 12. This increment translates into a larger number of basis states ($2^8= 256$ to $2^{12} = 4096$). Furthermore, the joint probability distribution that we encode in the target state is now three-dimensional by including an additional marginal distribution associated with the "forwardness" of the jet with respect to the beam (jet $\eta$). In Figure \ref{fig:3D_loss}, the JS divergence loss is plotted as a function of training step. The circuits were initialized in the all-zero state ($\ket{\psi_0}= \ket{0}^{\otimes 12}$) and $d=6(12)$ to construct our QCBM with Ansatz 1(2). In this plot, we can also see the effect of increasing the number of shots during training, reporting a significant difference in JS divergence values when increasing $N_{shots}$ from 8,192 to 20,000. 

\captionsetup[figure]{textfont=normalfont,justification=raggedright}

\begin{figure}
    \centering
    \includegraphics[width=\columnwidth]{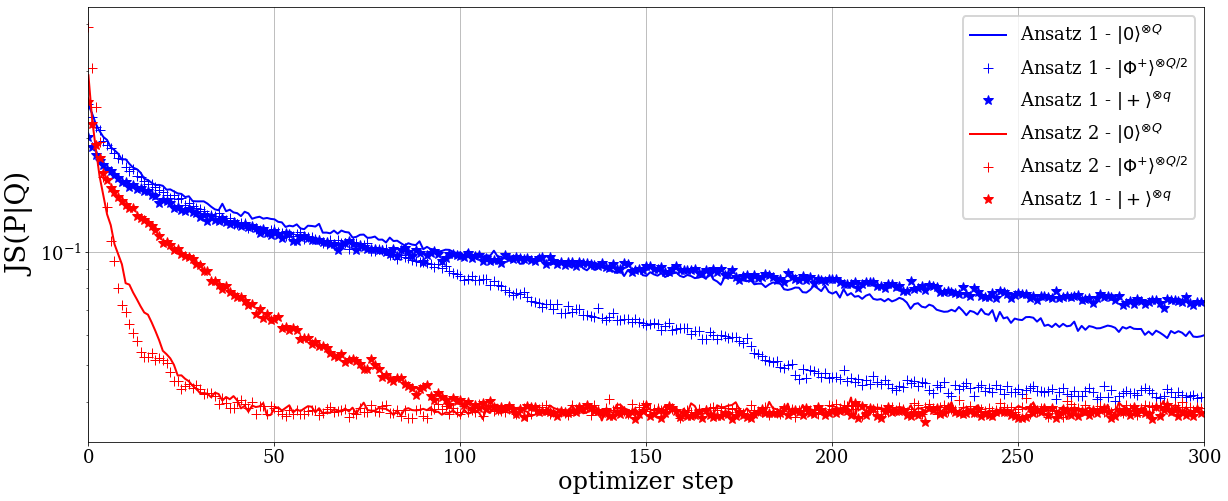}
    \caption{JS divergence as a function of training step using $N_{shots}=20,000$. Blue(red) points correspond to Ansatz 1(2). Circuits were initialized in the all-zero state (solid lines), a product of Bell states (cross), or the all-plus state (stars), and trained to learn a 3D joint distribution. 
    }
    \label{fig:3D_loss_initstate}
\end{figure}

When training QCBM with $Q=12$ qubits, we also used an initial state ($\ket{\psi_0}= \ket{\mathrm{GHZ}}^{\otimes Q/3}$) where $3$-qubit subsets are initialized in a GHZ state. The comparison for the JS divergence value as a function of the training step from the three initial states considered is displayed in Figure \ref{fig:3D_loss_initstate}. For Ansatz 1 (blue), the JS value for the three configurations is very similar for the first 100 training steps. From step 100 on, the QCBM initialized in a product of Bell states (blue cross) trains faster and converges to a lower JS value than the QCBMs initialized in the all-zero (solid blue) and all-plus (blue star) state. For Ansatz 2, the effect of the initial state used in the preparation of the circuit has a more negligible effect on the minimal JS value the model converges after training. Nonetheless, the QCBMs prepared from the all-plus state (stars) seem to take longer to converge. 

\captionsetup[figure]{textfont=normalfont,justification=raggedright}

\captionsetup[figure]{textfont=normalfont,justification=raggedright}
\begin{figure}
    \centering
    \includegraphics[width=\columnwidth]{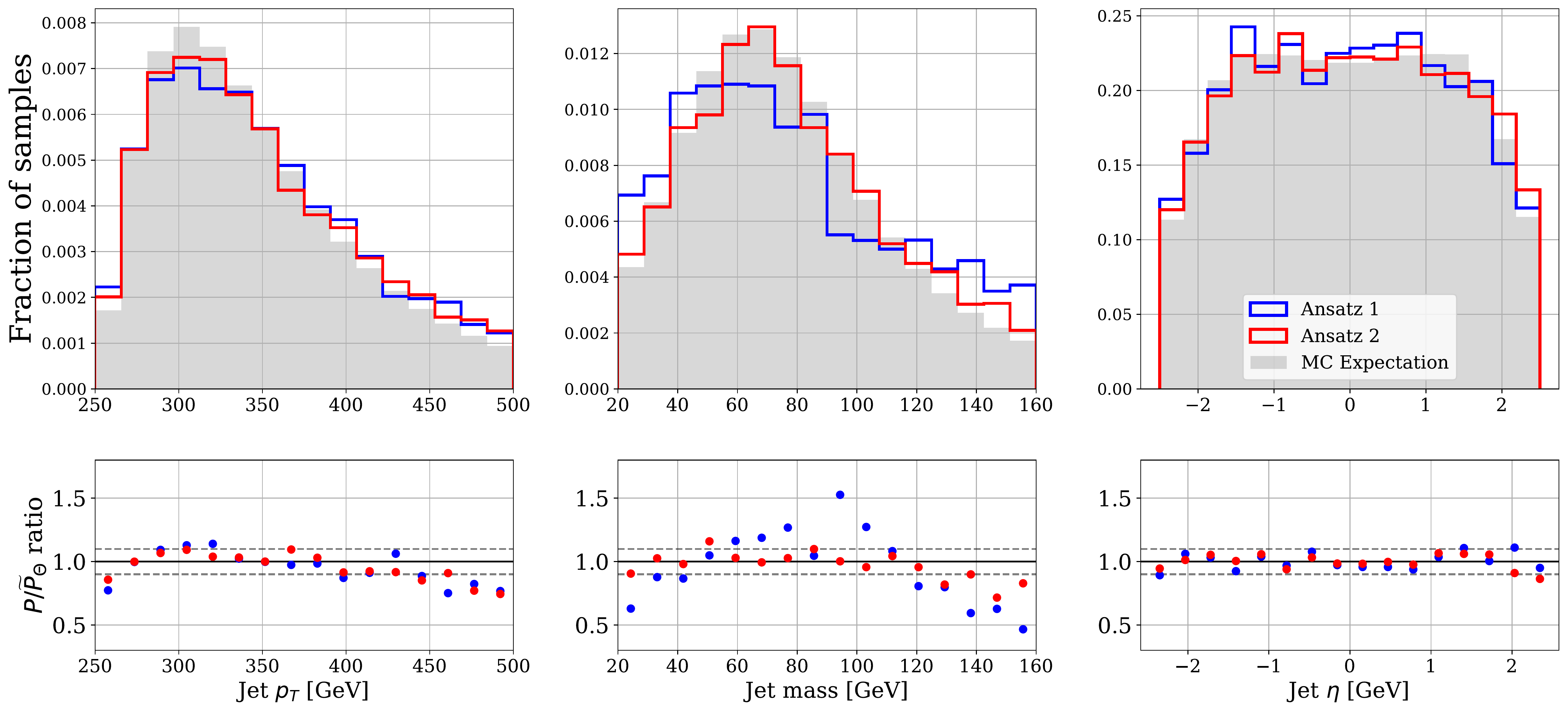}\\
    \caption{\textbf{Top:} Target and sampled distributions. Blue(red) points correspond to Ansatz 1(2). Circuits were initialized in the all zero state ($\ket{\psi_0}= \ket{0}^{\otimes 12}$) and trained to learn a 3D joint distribution. \textbf{Bottom:} Ratio of target and sampled distributions with horizontal guide lines marking $P/\widetilde{P}=0.9$ and $P/\widetilde{P}=1.1$.}
    \label{fig:3D_allzero}
\end{figure}

\captionsetup[figure]{textfont=normalfont,justification=raggedright}
\begin{figure}
    \centering
    \includegraphics[width=\columnwidth]{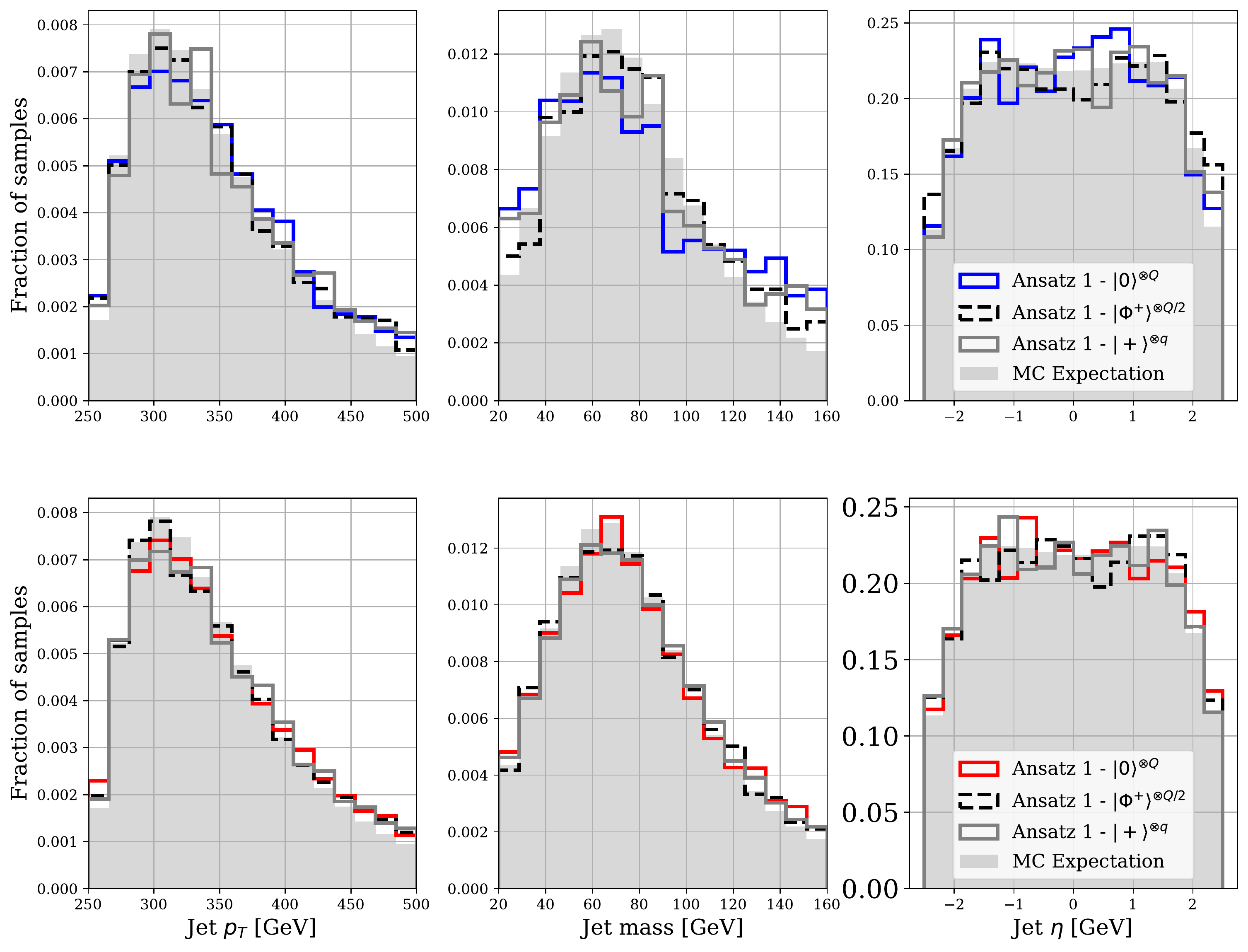}\\
    \caption{\textbf{Top:} Target and sampled distributions for 12-qubit QCBM  trained to learn a 3D joint distribution using Ansatz 1. Registers were initialized in the all zero state (solid blue line), a product of Bell states (black dashed line), and the all-plus state (gray dashed line). 
    \textbf{Bottom:} Target and sampled distributions for 12-qubit QCBM  trained to learn a 3D joint distribution using  Ansatz 2. Registers were initialized in the all zero state (solid red line), a product of Bell states (black dashed line), and the all-plus state (gray dashed line).}
    \label{fig:3D_init_state}
\end{figure}

\begin{table}[h!]
\renewcommand{\arraystretch}{1.5}
\centering
 \begin{tabular}{|c| c| c| c|} 
 \hline
 Ansatz & $N_{\rm par}$ & Initial State & $D(p|\widetilde{p}(\Theta))$  \\ [0.5ex] 
 \hline\hline
 1 & \multirow{3}{*}{432} &$\ket{\psi_0}= \ket{0}^{\otimes 12}$ & 0.0226 \\ 
 1 & &$\ket{\Phi^+}^{\otimes 6}$     & 0.0138 \\
 1 & &$\ket{\psi_0}= \ket{\mathrm{GHZ}}^{\otimes 4}$ & 0.0187 \\
 \hline
 2 & \multirow{3}{*}{468} &$\ket{\psi_0}= \ket{0}^{\otimes 12}$ & 0.0108 \\
 2 & &$\ket{\Phi^+}^{\otimes 6}$     & 0.0090 \\
 2 & &$\ket{\psi_0}= \ket{\mathrm{GHZ}}^{\otimes 4}$& 0.0106\\[1ex]
 \hline
 \end{tabular}
 \caption{Comparison between target and sampled distributions according to Eq. \ref{eq:dif}.}
 \label{table:3dcomp}
\end{table}

In Figure \ref{fig:3D_allzero}, we compare the distributions of samples generated via projective measurements on the circuits evaluated on the trained parameters. The circuits were initialized in the all-zero state. By looking at Figure \ref{fig:3D_allzero} and Table \ref{table:3dcomp}, we observe a degraded performance in terms of similarity metric (Eq. \ref{eq:dif}) when compared to the 8-qubit circuit results. The $D(p|\widetilde{p}(\Theta))$ value increases from 0.007153 to 0.02226 for Ansatz 1 and from 0.005452 to 0.0108 for Ansatz 2. In Figure \ref{fig:3D_init_state}, we compare the distributions generated by QCBMs initialized in the three different initial configurations. Again, the trained QCBM that prepares the target distribution with the highest fidelity is Ansatz 2 and we see little dependence on the initial state for QCBMs prepared using Ansatz 2. Nonetheless, the effect of the initial state in QCBMs prepared using Ansatz 1 is now more evident.  

\begin{table}[H]
\centering
\scriptsize
\begin{subtable}[t]{0.4\textwidth}
\centering 
\caption{Monte Carlo (Ground Truth)}
        \label{tab:MC_3d_correlation}
\begin{tabular}{|p{1cm}|c|c|c|}
\cline{2-4}
\multicolumn{1}{c|}{} & $p_{T}$ & mass  & $\eta$ \\
\hline
 $p_{T}$ & {\cellcolor{gray!20}-} & 0.2 & 7.3e-12 \\ \hline
 mass & 0.2 & {\cellcolor{gray!20}-} & 2.7e-11 \\ \hline
 $\eta$ & 7.3e-12 & 2.7e-11 &  {\cellcolor{gray!20}-} \\ \hline
\end{tabular}\\
\end{subtable}
\begin{subtable}[t]{0.5\textwidth}
\centering
\caption{Ansatz 1}
\label{tab:ansatz1_3d_correlation}
\resizebox{1\textwidth}{!}{
 \begin{tabular}{| p{0.25cm} | c | c | c |c |c |c | c| c| c|}
\cline{2-10}
 \multicolumn{1}{c|}{} & \multicolumn{3}{c|}{$p_{T}$} & \multicolumn{3}{c|}{mass (m)}  & \multicolumn{3}{c|}{$\eta$}\\ 
\cline{2-10}
 \multicolumn{1}{c|}{} & $\ket{0}^{\otimes 12}$  & $\ket{\Phi^+}^{\otimes 6}$ & $\ket{\mathrm{GHZ}}^{\otimes 4}$ & $ \ket{0}^{\otimes 12}$  & $\ket{\Phi^+}^{\otimes 6}$ & $\ket{\mathrm{GHZ}}^{\otimes 4}$ & $\ket{0}^{\otimes 12}$  & $\ket{\Phi^+}^{\otimes 6}$ & $ \ket{\mathrm{GHZ}}^{\otimes 4}$\\ [0.5ex] 
 \hline
 $p_{T}$ & \multicolumn{3}{c|}{\cellcolor{gray!20}-} & {\cellcolor{olive!20} 0.16} & {\cellcolor{teal!20} 0.16}& {\cellcolor{violet!20}0.18} & {\cellcolor{olive!20} 8.2e-3} & {\cellcolor{teal!20} 1.2e-3} & {\cellcolor{violet!20}-1.3e-3}  \\
 \hline
 m    & {\cellcolor{olive!20} 0.16} & {\cellcolor{teal!20} 0.16} & {\cellcolor{violet!20}0.18} & \multicolumn{3}{c|}{\cellcolor{gray!20}-}  & {\cellcolor{olive!20} -0.014} & {\cellcolor{teal!20} 4.4e-3} & {\cellcolor{violet!20}7.7e-3}\\
 \hline
 $\eta$  & {\cellcolor{olive!20} 8.2e-3} & {\cellcolor{teal!20} 1.2e-3} & {\cellcolor{violet!20}-1.3e-3} & {\cellcolor{olive!20} -0.014} & {\cellcolor{teal!20} 4.4e-3} & {\cellcolor{violet!20}7.7e-3} & \multicolumn{3}{c|}{\cellcolor{gray!20}-} \\
 \hline
 \end{tabular}
 }
 \end{subtable}
 \begin{subtable}[t]{0.5\textwidth}
\centering
\caption{Ansatz 2}
\label{tab:ansatz2_3d_correlation}
\resizebox{1\textwidth}{!}{
 \begin{tabular}{| p{0.25cm} | c | c | c |c |c |c | c| c| c|}
\cline{2-10}
 \multicolumn{1}{c|}{} & \multicolumn{3}{c|}{$p_{T}$} & \multicolumn{3}{c|}{mass (m)}  & \multicolumn{3}{c|}{$\eta$}\\ 
\cline{2-10}
 \multicolumn{1}{c|}{} & $\ket{0}^{\otimes 12}$  & $\ket{\Phi^+}^{\otimes 6}$ & $\ket{\mathrm{GHZ}}^{\otimes 4}$ & $ \ket{0}^{\otimes 12}$  & $\ket{\Phi^+}^{\otimes 6}$ & $\ket{\mathrm{GHZ}}^{\otimes 4}$ & $\ket{0}^{\otimes 12}$  & $\ket{\Phi^+}^{\otimes 6}$ & $ \ket{\mathrm{GHZ}}^{\otimes 4}$\\ [0.5ex] 
 \hline
 $p_{T}$ & \multicolumn{3}{c|}{\cellcolor{gray!20}-} & {\cellcolor{olive!20} -4.1e-3} & {\cellcolor{teal!20} 4.6e-3}& {\cellcolor{violet!20}0.019} & {\cellcolor{olive!20} -3.7e-3} & {\cellcolor{teal!20} -9.1e-3} & {\cellcolor{violet!20}1.6e-3}  \\
 \hline
 m    & {\cellcolor{olive!20} -4.1e-3} & {\cellcolor{teal!20} 4.6e-3} & {\cellcolor{violet!20}0.019} & \multicolumn{3}{c|}{\cellcolor{gray!20}-}  & {\cellcolor{olive!20} 6.3e-3} & {\cellcolor{teal!20} 4.9e-3} & {\cellcolor{violet!20}2.2e-3}\\
 \hline
 $\eta$  & {\cellcolor{olive!20} -3.7e-3} & {\cellcolor{teal!20} -9.1e-3} & {\cellcolor{violet!20}1.6e-3} & {\cellcolor{olive!20} 6.3e-3} & {\cellcolor{teal!20} 4.9e-3} & {\cellcolor{violet!20}2.2e-3} & \multicolumn{3}{c|}{\cellcolor{gray!20}-} \\
 \hline
 \end{tabular}
 }
 \end{subtable}
 \caption{Correlation matrices between jet $p_{T}$, mass (m), and $\eta$ variables in the (a) target distribution, and samples obtained from the evaluation of the QCBMs constructed using (b) Ansatz 1 and (c) Ansatz 2 in Figure \ref{fig:circuid_diagram} with the trained parameters. Values displayed for initial states prepared in the all-zero state (olive), a product of 6 Bell states (teal), and a product of 4 GHZ states (violet).}
 \label{table:3dcorrMatrix}
\end{table}

Finally, we report on the correlation matrix between the jet $p_{T}$, mass and $\eta$ variables used to construct the target distribution. In Table \ref{table:3dcorrMatrix}, the values associated with the target distribution are displayed in black. The results in Table \ref{table:3dcorrMatrix} display a slight discrepancy in the original correlation matrix (training dataset) and that for the samples generated by sampling Ansatz 1 (blue) with the trained parameters. Again, for the QCBM constructed with Ansatz 2, the matrix values indicate that there is little correlation between the marginal distributions in the synthetic samples.

\subsection{Noisy Training with Layer-wise Coordinate Descent}
\label{sec:LCD_training}
The gradient-based training of the QCBM models presented in \ref{sec:training} was executed on noiseless (ideal) qubits but was not used for training on noisy hardware.   However, the performance of the trained QCBM model on near-term quantum devices will be heavily impacted by hardware noise.  Qubit initialization, gate noise, and errors in the circuit measurement step, all result in state preparation error that can cause parameterized models to converge to maximally mixed states, with an overall effect of flattening the loss landscape \cite{wang2020noiseinduced} and manifests in a noisy estimation of $\widetilde{P}_{\Theta}(x)$. 

While developing error mitigation methods that can be incorporated into variational training algorithms is an open area of research, one approach to noise mitigation is to implement the circuit training with hardware noise in order to learn optimized parameters that can compensate for time-independent errors, such as over- and under-rotation in single qubit gates.  We test the robustness of the final parameters found in Section \ref{sec:training} to hardware noise by first executing the trained QCBM models constructed with Ansatz 2 on superconducting qubit devices. The $8$-qubit QCBM (276/312 parameters) was executed on the $16$-qubit device \texttt{ibmq\_guadalupe} and the $12$-qubit QCBM (468) was executed on the $27$-qubit device \texttt{ibm\_cairo}, both were accessed through a cloud-based queue. 

\begin{figure*}
    \centering
    \includegraphics[width=\textwidth]{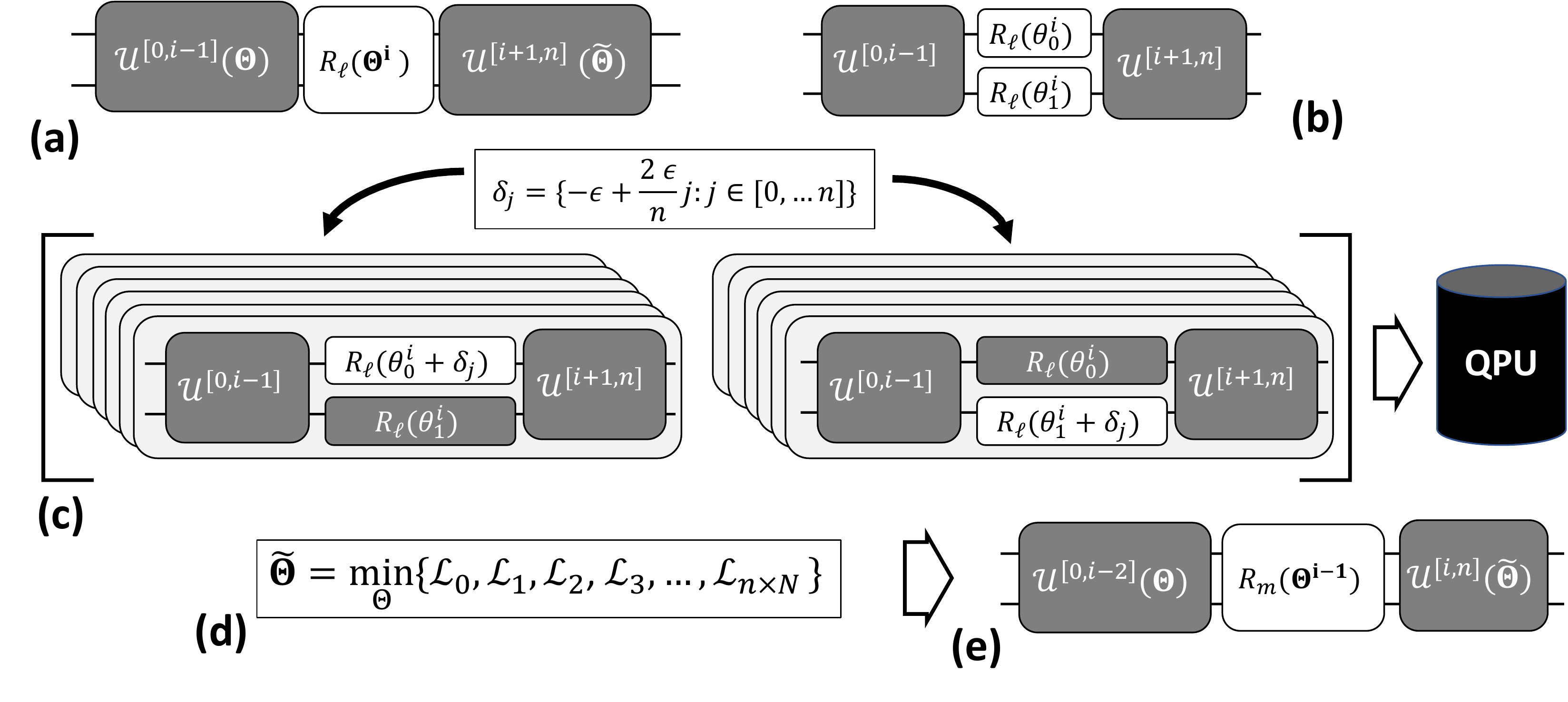}
    \caption{Layer-wise coordinate descent (LCD) workflow.  (a) a single rotational layer in a $N$-qubit QCBM model, (b) is composed of rotational gates $R_{\ell}$ acting on individual qubits. A batch of $n \times N$ circuits is constructed by sweeping each individual gate over a discrete set of $n$ shifts and executed on a quantum processor (QPU).  The loss is evaluated for each circuit executed (d), the parameter vector $\Theta$ is updated by the values which return the minimal value, and the updated parameter vector is used to start the search over the next rotational layer (e).}
    \label{fig:layerwise_tuning_workflow}
\end{figure*}

In the absence of error and noise mitigation we instead opted to use a localized search over individual parameters.  Each parameterized Ansatz (shown in Fig. \ref{fig:circuid_diagram}) are constructed with layers of parameterized rotation gates, implemented using an arbitrary unitary gate with $3$ rotational parameters. Then, we used a layer-wise coordinate descent (LCD) method to optimize each QCBM performance on hardware.  The workflow is shown in Fig. \ref{fig:layerwise_tuning_workflow}.  No readout error mitigation or other noise mitigation methods were used.  
\captionsetup[figure]{textfont=normalfont,justification=raggedright}
LCD optimizes the parameters of a circuit $\mathcal{U}(\Theta)$ by searching the multi-dimensional parameter space along linear cuts of finite width. With $N$ qubits in the register, each parameter was swept through a shift of parameters defined by $\epsilon=-\pi/4$ and spacing $2\epsilon/n$. The targeted backends allowed for a maximum number of circuits per batch ($B$) which defines the spacing $n = B/N$.  For the $8$-qubit QCBM trained on \texttt{ibmq\_guadalupe} this resulted in a grid spacing of $0.0419$.  For the $12$-qubit QCBM trained on \texttt{ibm\_cairo} this resulted in a mesh spacing of $0.0628$. Each circuit was sampled using $N_{shots} = 20 000$.The rotational parameters are optimized starting with the gates closes to the measurement process. If the noiseless simulation returned an excessively flat landscape (variance of the loss was below $5e-7$) these experiments were not executed on hardware. 

\begin{figure*}
    \centering
    \includegraphics[width=\textwidth]{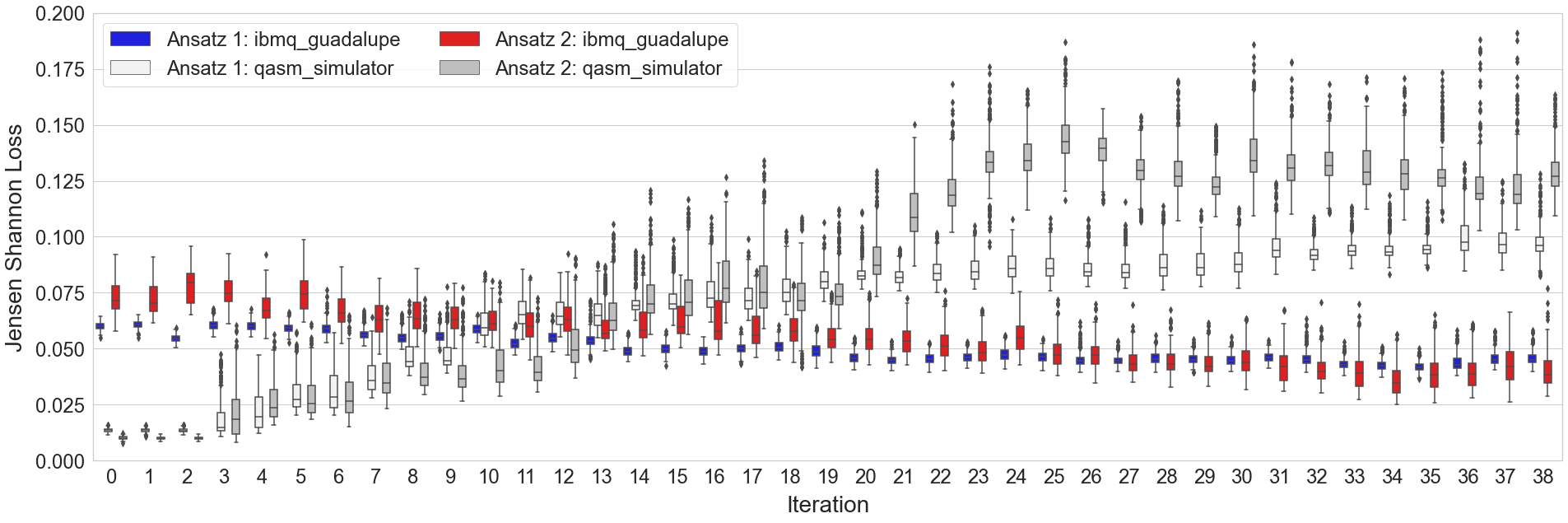}\\
    \includegraphics[width=\textwidth]{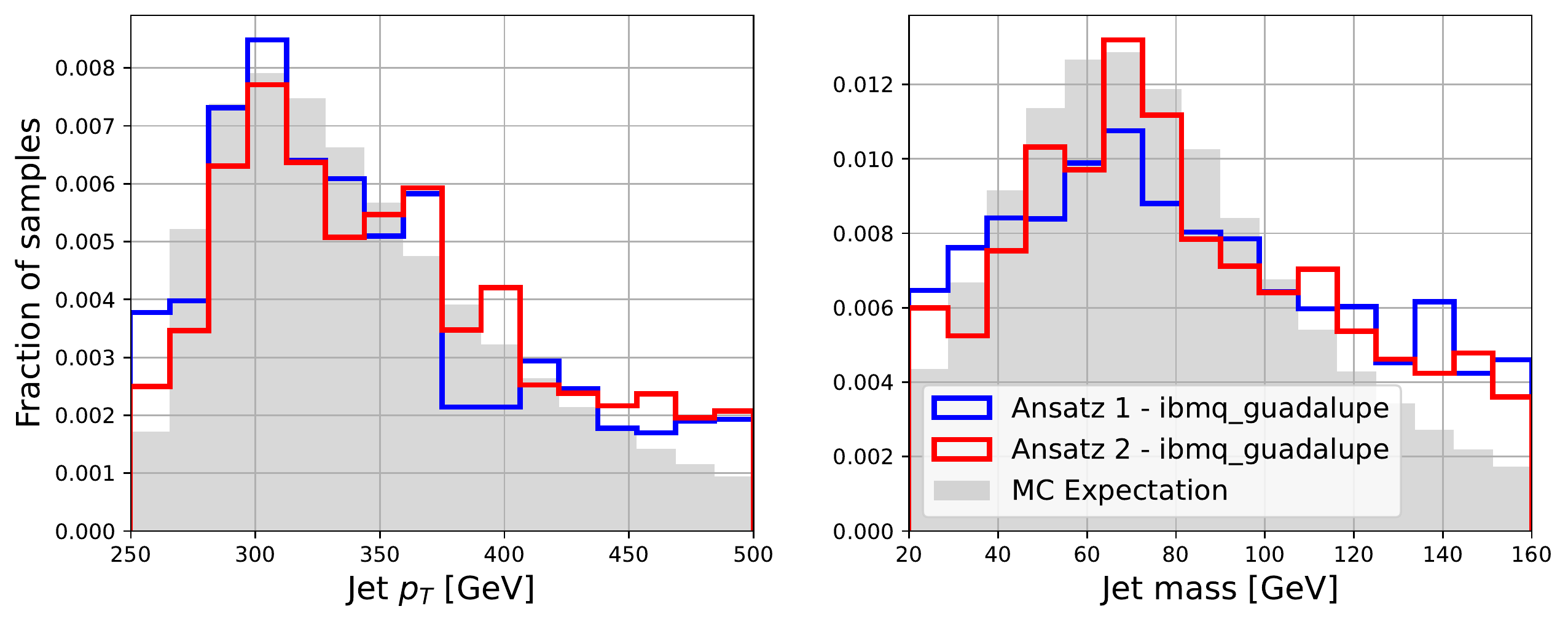}
    \caption{\textbf{Top:} Quartiles of Jensen-Shannon loss over each iteration of LCD with 8 qubit QCBM circuits. \textbf{Bottom:} Target and sampled distributions constructed using Ansatz 1(2) in blue (red). Circuits were initialized in the all-zero state and evaluated on \texttt{ibmq\_guadalupe} with an $8$-qubit QCBM. Distributions for the best parameters found after the execution of the LCD workflow.}
    \label{fig:layerwise_tuning_8qubits}
\end{figure*}

\begin{figure*}
    \centering
    \includegraphics[width=\textwidth]{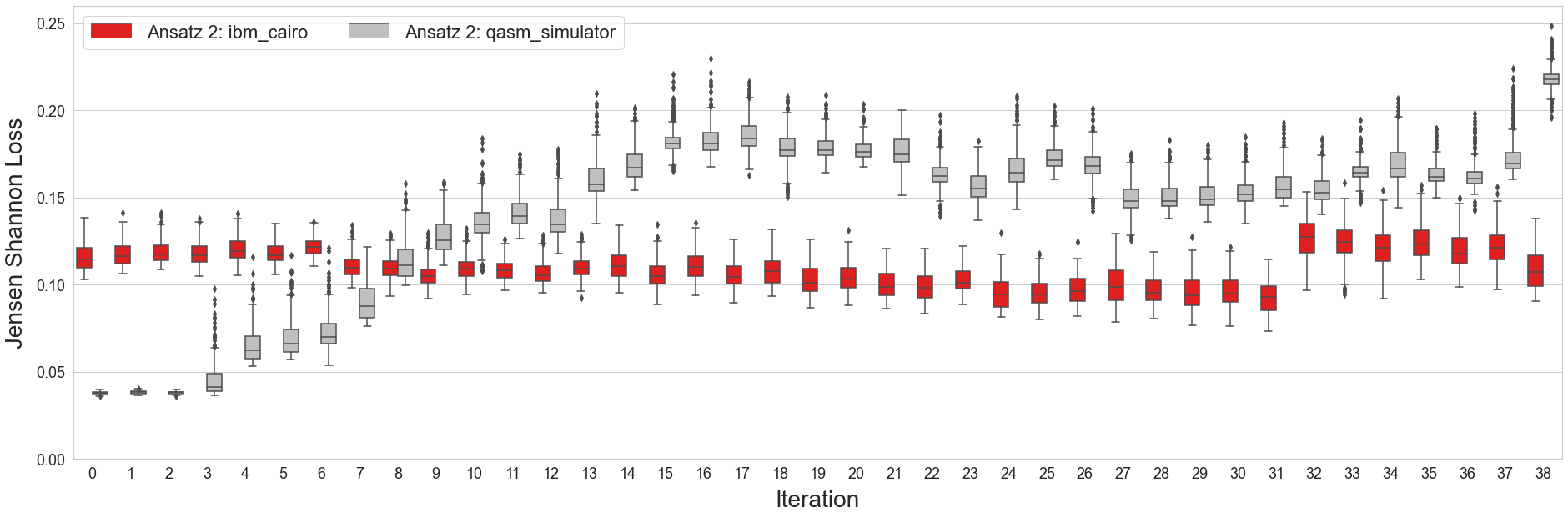}\\
    \includegraphics[width=\textwidth]{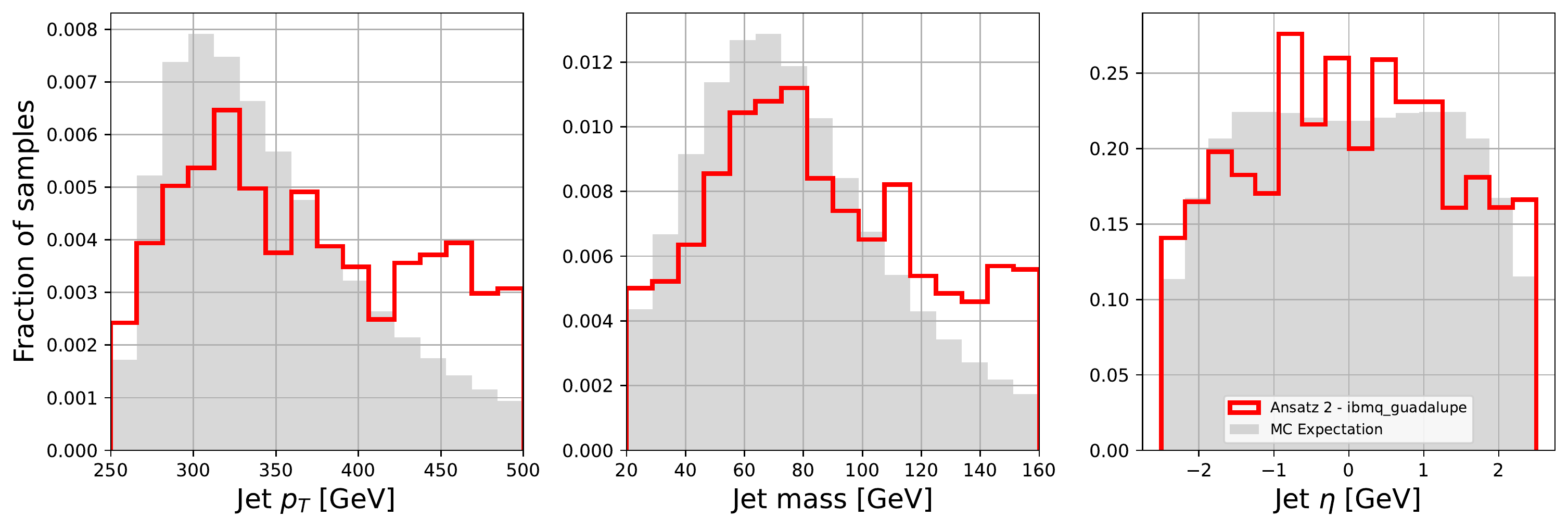}
    \caption{\textbf{Top:} Quartiles of Jensen-Shannon loss over each iteration of LCD implemented in \texttt{ibm\_cairo} with a $12$-qubit QCBM. \textbf{Bottom:} Target (gray) and sampled distributions constructed using Ansatz 2 in blue (red lines). Circuits were initialized in the all-zero state and evaluated on \texttt{ibm\_cairo} with an $12$-qubit QCBM. Distributions for the best parameters found after the execution of the LCD workflow.} 
    \label{fig:layerwise_tuning_12qubits}
\end{figure*}

The top of Figure \ref{fig:layerwise_tuning_8qubits} displays the quartiles of JS loss over each iteration of LCD with 8-qubit QCBM circuits. The quartiles are plotted for the QCBM constructed with Ansatz 1 and 2 and evaluated with the updated parameters after each iteration on the \texttt{ibmq\_guadalupe} backend in blue and red, respectively. The plot also shows how the JS divergence value degrades as the LCD training parameters deviate from those obtained during the noiseless training. On the other hand, hardware performance is relatively stable. The bottom of Figure \ref{fig:layerwise_tuning_8qubits} shows the sampled distributions generated by the evaluation of the QCBM with the parameters that yielded the lowest JS divergence value during the LCD training. We observe a more significant discrepancy between the target and sampled distributions compared to the results reported in Figure \ref{fig:2D_allzero}, where the QCBM is evaluated with the parameters obtained during the noiseless training. In Figure \ref{fig:layerwise_tuning_12qubits}, the quartiles of JS loss (top) and the sampled and target distributions (bottom) are displayed. The QCBMs are constructed using Ansatz 2, to prepare a 3D joint distribution in a 12-qubit register. The LCD training was performed on the \texttt{ibm\_cairo} device. 

\captionsetup[figure]{textfont=normalfont,justification=raggedright}

\captionsetup[figure]{textfont=normalfont,justification=raggedright}

\begin{figure*}
    \centering
    \includegraphics[width=0.92\linewidth]{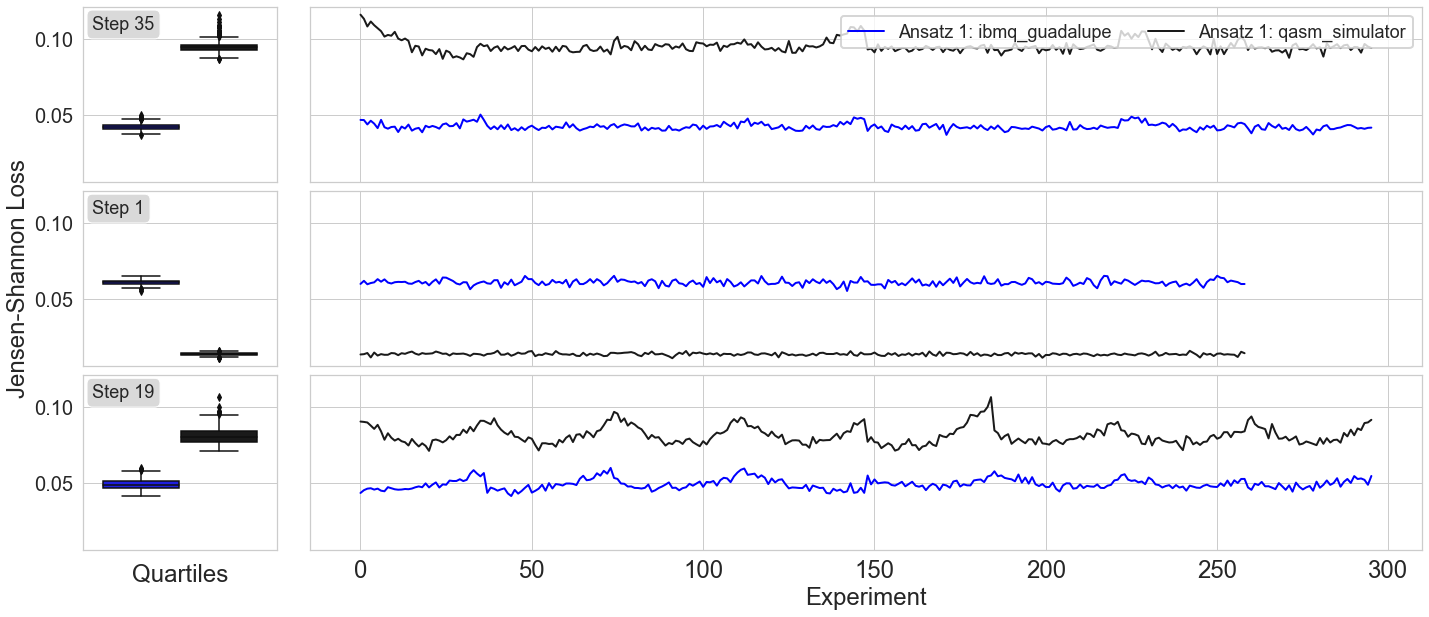}\\
    \includegraphics[width=0.92\linewidth]{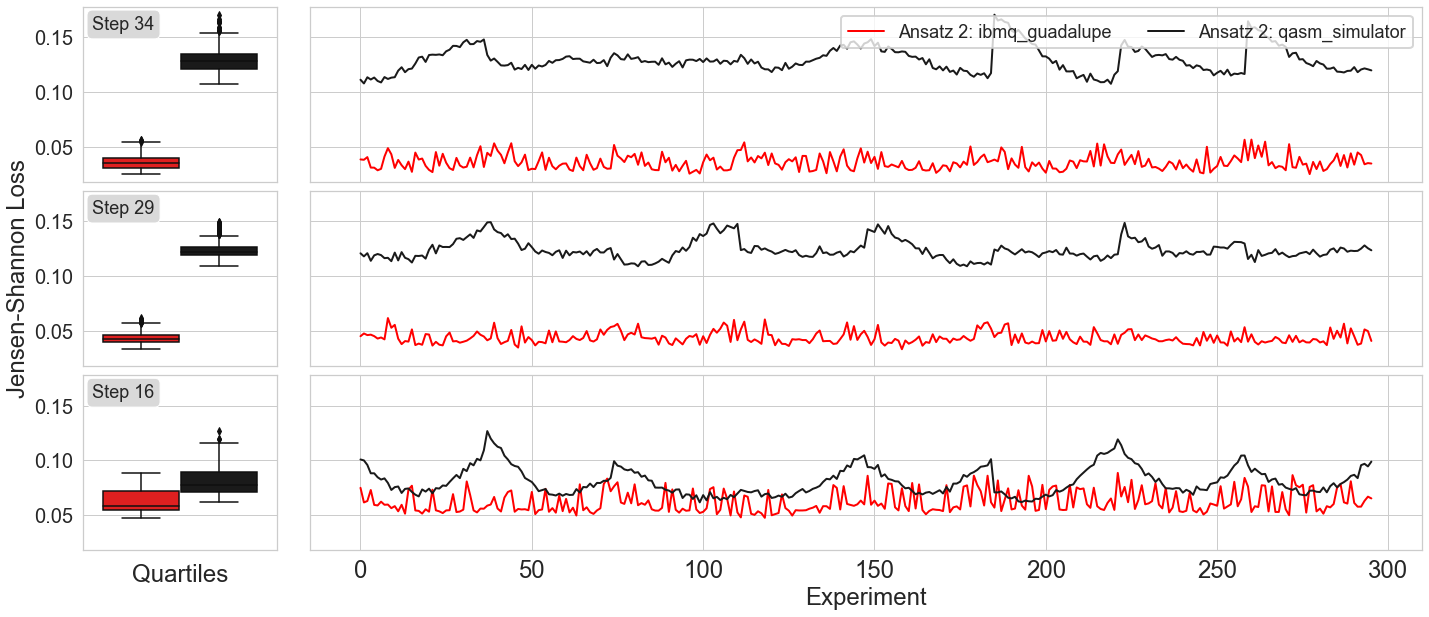}\\
     \includegraphics[width=0.92\linewidth]{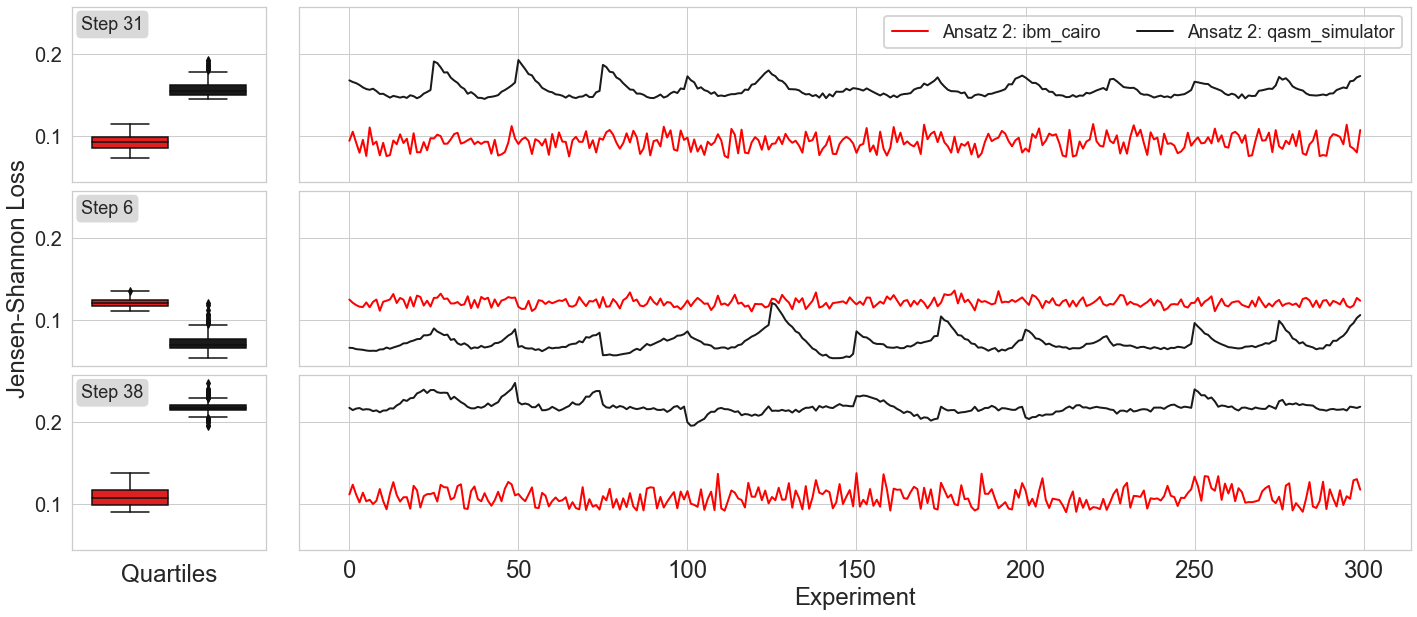}
    \caption{\textbf{Left:} Quartiles of Jensen-Shannon loss for three representative iterations in the LCD scheme. Each iteration \textbf{right} corresponds to a parameter sweep in the range $[-\frac{\pi}{4}, \frac{\pi}{4}]$ for a particular rotation gate in a given layer. The plots display the JS values for a 8-qubit QCBM trained on \texttt{ibmq\_guadalupe} using Ansatz 1 (top) and Ansatz 2(middle). The bottom plot corresponds to a 12-qubit QCBM constructed using Ansatz 2 and trained on \texttt{ibm\_cairo}. }
    \label{fig:layerwise_tuning_slices_quartiles}
\end{figure*}

In Figure \ref{fig:layerwise_tuning_slices_quartiles} we show the quartiles of JS for three iterations representative of the different stages of the LCD training (left). During the first iterations in the training, the loss landscape is very flat, and the JS value is lower for the QCBMs evaluated on the \texttt{qasm\_simulator}. The box plots on the left correspond to the JS values obtained during the parameter sweep in the range $\epsilon \in [-\frac{\pi}{4}, \frac{\pi}{4}]$ for a particular rotation gate in a given layer. Then, we observe that, for subsequent iterations, the parameter tuning moves the parameters into regions where the performance of the noiseless simulator is degraded, and slowly improving the performance on the quantum device.

As we can see from Figures \ref{fig:layerwise_tuning_8qubits} and \ref{fig:layerwise_tuning_12qubits}, noisy training can improve performance, but there are a number of obstacles.  When the training is initialized with parameters pre-trained with noiseless qubits, multiple interations may be needed to re-train.  The gradual improvements in the loss function may not be robust against large deviations in the device noise.  For example, executing the LCD workflow over all rotational parameters of the 12-qubit QCBM was done over multiple days, probably causing the discrete change in the loss between iterations $31$ and $32$. 

\section{QCBM Design Space}
\label{sec:design_space}
Parameterized circuit ansatz used for variational algorithms need to balance expressability with trainability.  There are considerable ongoing efforts in determining the characteristics of circuits that lead to effective training and scaling. In this paper we used two different ansatz designs and multiple initializations for the $Q$-qubit register.  In this section we discuss some observations based on our results reported in Sections \ref{sec:training} and \ref{sec:LCD_training}.  The QCBM models we trained in this paper used the same parameterization (the arbitrary rotation gate implemented in PennyLane) and entangling gate operation (the CNOT gate).  Each feature was encoded into the same number of qubits ($4$) which defined the size of the overall register: $8$ qubits for 2D distributions, $12$ qubits for 3D distributions. 

Between the two ansatz designs, only Ansatz 1 can generate arbitrary $Q$-qubit entanglement and fit arbitrary correlations between 2- or 3-variables, regardless of the choice of initial state.  However, as shown in Figs. \ref{fig:2D_loss},\ref{fig:2D_all},\ref{fig:3D_loss},\ref{fig:3D_init_state} this ansatz slowly learns. On the other hand, Ansatz 2 quickly learns, but only prepares a product state of 4-qubit systems.  When the qubit register is initialized in the all zero state $|0\rangle^{\otimes Q}$ or as a set of Bell states $|\Phi^+\rangle^{\otimes Q}$, then Ansatz 2 cannot by definition, model arbitrary correlations between each 4 qubit subset.  For $Q=12$, if the circuit is initialized with $|\mathrm{GHZ}\rangle^{\otimes 4}$, then there is local entanglement between the qubit subsets.  Yet as reported in Tables \ref{table:2dcorrMatrix} and \ref{table:3dcorrMatrix},  this is insufficient to capture the correlations as seen in the Monte Carlo data.  We observe a trade-off in the modeling capacity of Ansatz 1 and Ansatz 2: Ansatz 1 can model the correlations between variables but has lower fidelity in fitting marginal distributions (as quantified by Eq. \ref{eq:dif} in Tables \ref{table:2dcomp} and \ref{table:3dcomp} and seen in Figs. \ref{fig:2D_allzero},\ref{fig:3D_allzero}). On the other hand, for Ansatz 2 the generated data fails to capture the correlations in the Monte Carlo data, but has high fidelity in fitting marginal distributions (as reported in Tables \ref{table:2dcorrMatrix},\ref{table:3dcorrMatrix}). Simply including local correlations in the initial state  by using $\ket{\rm GHZ}^{\otimes 4}$ was insufficient to generate high correlations between variables $p_T$ and mass. 

\section{Conclusion}
The size of the design space associated with parameterized quantum circuit models is large. This work demonstrated the efficacy of non-adversarial unsupervised training of generative models implemented as parameterized quantum circuits. We demonstrate the usefulness of these quantum models in the context of a HEP application and to assist other practitioners, we have used: several circuit ansatzes found in the quantum computing literature and tested the trainability of QCBM initialized with different quantum states.  We quantify the fidelity of the trained models using: the JS score (also used to train the models), the MAE of feature marginals, and the correlation matrices of generated data. 

We are encouraged by the success of gradient-based training for 12 qubit QCBM. We show that for two and three correlated variables, both ansatz can minimize the loss to the order of $\sim 10^{-2}$, but whether that corresponds to models that can faithfully reproduce the kinematic distributions of a jet in a $pp$ collision typical of the LHC experiment cannot be deduced by the training loss alone. Only Ansatz 1 can fit the correlations between variables in the absence of noise, but the individual marginal fits for Ansatz 1 are lower fidelity than Ansatz 2. On the other hand, while Ansatz 2 can reproduce the individual marginals with high fidelity it cannot by definition, model correlations, as we see in Tables \ref{table:2dcorrMatrix} and \ref{table:3dcorrMatrix}. We report these results to assist other practitioners in the design of parameterized ansatz for scientific applications.

We also report on the influence of hardware noise on the QCBM performance. In our study, the QCBM were trained in the absence of noise and certain trained models were deployed on near-term devices. In our results we observe that hardware noise flattens out the landscape.  Additionally, we observe that the addition of hardware noise does not lead to an increase in correlation between the encoded variables.

Training in the presence of hardware noise (e.g. using local parameter tuning or LCD) can improve performance, however, without robust error mitigation hardware fluctuations can undo small improvements in performance (see Fig. \ref{fig:layerwise_tuning_12qubits}). Designing scale-able error mitigation methods that fully capture correlations in the hardware is an active area of research in quantum computing. For example, commonly employed methods of readout error mitigation using measurement fidelity matrices \cite{hamilton2019errormitigated,hamilton2020scalable}. These methods have an advantage in that the can be incorporated in to variational training workflows (either gradient-based optimization or LCD) as a data post-processing step. This motivates the need for a systematic follow up study of error mitigation efficacy in this application, with a focus on balancing overhead with model performance. 

\begin{acknowledgments}
This work was partially supported by the Quantum Information Science Enabled Discovery (QuantISED) for High Energy Physics program at ORNL under FWP ERKAP61. This work was partially supported by the Laboratory Directed Research and Development Program of Oak Ridge National Laboratory, managed by UT-Battelle, LLC, for the U. S. Department of Energy. This work was partially supported as part of the ASCR Testbed Pathfinder Program at Oak Ridge National Laboratory under FWP ERKJ332.  This work was partially supported as part of the ASCR Fundamental Algorithmic Research for Quantum Computing Program at Oak Ridge National Laboratory under FWP ERKJ354. This research used quantum computing system resources of the Oak Ridge Leadership Computing Facility, which is a DOE Office of Science User Facility supported under Contract DE-AC05-00OR22725. Oak Ridge National Laboratory manages access to the IBM Q System as part of the IBM Q Network. The authors would like to thank Dr. Phil Lotshaw for helpful comments during the manuscript preparation.  

\end{acknowledgments}

\bibliography{ref}

\end{document}